\documentclass[10pt]{amsart}
\usepackage{enumerate,amsmath,amssymb,latexsym,
amsfonts, amsthm, amscd}

\usepackage{enumerate,amsmath,amsthm,amssymb,latexsym,
amsfonts,amscd}
\usepackage[all]{xy}

\title[Covariant energy density of the gravitational field]{The fully covariant energy\\ momentum stress tensor
 of the gravitational field and \\
the Einstein equation for gravity in general relativity\\}

\author[Maurice J. Dupr\'e]{M\lowercase{aurice} J. D\lowercase{upr\'e}\\D\lowercase{epartment of} M\lowercase{athematics}
\\N\lowercase{ew} O\lowercase{rleans}, LA 70118\\\lowercase{email:  mdupre@tulane.edu}\\4 M\lowercase{arch} 2009}
\address{DEPARTMENT OF MATHEMATICS\\TULANE UNIVERSTIY\\NEW ORLEANS, LA 70118}
\email{mdupre@tulane.edu}

\theoremstyle{plain}

\newtheorem{proposition}{Proposition}[section]
\newtheorem{theorem}{Theorem}[section]
\newtheorem{corollary}{Corollary}[section]
\newtheorem{postulate}{Postulate}[section]

\theoremstyle{definition}

\numberwithin{equation}{section}

\newcommand{\B}{\mathcal B}
\newcommand{\Ce}{\mathcal C}
\newcommand{\D}{\mathcal D}

\newcommand{\F}{\mathcal F}

\newcommand{\K}{\mathcal K}

\newcommand{\M}{\mathcal M}

\newcommand{\R}{\mathcal R}
\newcommand{\T}{\mathcal T}

\newcommand{\bC}{\mathbb{C}}

\newcommand{\bK}{\mathbb{K}}
\newcommand{\bN}{\mathbb{N}}

\newcommand{\bR}{\mathbb{R}}

\newcommand{\ra}{\rightarrow}

\newcommand{\lra}{\longrightarrow}

\newcommand{\del}{\partial}

\newcommand{\med}{\medbreak}

\begin{document}

\maketitle

\begin{abstract}
We give a fully covariant energy momentum stress tensor for the
gravitational field which is easily physically and intuitively
motivated, and which leads to a very general derivation of the
Einstein equation for gravity. We do not need to assume any
property of the source matter fields' energy momentum stress
tensor other than symmetry. We give a physical motivation for this
choice using laser light pressure. As a consequence of our
derivation, the energy momentum stress tensor for the total source
matter fields must be divergence free, when spacetime is 4
dimensional. Moreover, if the total source matter fields are
assumed to be divergence free, then either the spacetime is of
dimension 4 or the spacetime has constant scalar curvature.
\end{abstract}


\med \textbf{Mathematics Subject Classification (2000)} : 83C05,
83C40, 83C99.

\med \textbf{Keywords} : Gravity, general relativity, Einstein
equation, energy density.

\section{INTRODUCTION}

\med

Our purpose here is two-fold.  First, we wish to give a fully
covariant energy momentum stress tensor for the gravitational
field.  Second, we will use our gravitational field energy
momentum stress tensor to give a general derivation of the
Einstein equation for gravity, and find as a consequence that the
divergence of the energy momentum stress tensor for matter and
fields other than gravity must be zero.

This manuscript is an expanded version of a manuscript submitted
for publication.  In communication with physicists, it has come to
my attention that in general, they are not as comfortable as
mathematicians with some of the more modern results in analysis
and differential topology and geometry. Here we will attach an
appendix for each mathematical topic that possibly needs more
coverage.

Since Einstein's and Hilbert's original "derivations" of the
Einstein equation for gravity in classical general relativity
(CGR), there have appeared too many different derivations to list.
The many different types of derivations are summarized in
\cite{SACHSWU}. In fact, all these subsequent derivations as well
as Hilbert's original derivation contrast markedly from Einstein's
original derivation and usually appeal to some abstract
mathematical principle which though desirable, is usually not
justifiable beyond mere desire. For instance, one of the most
popular textbook derivations simply modifies one side of the
equation to make it have zero divergence on grounds that physical
considerations make the other side, the matter energy momentum
stress tensor, have zero divergence.

If one uses a Lagrangian or variational method, then one is
immediately faced with the question of justifying the choice of
Lagrangian which is generally not really possible. In fact, to
quote from \cite{KRIELE}, after a detailed rigorous treatment of
the Lagrangian formulation of Einstein's equation (pages 271-279),
"That the Lagrangian ansatz described in this section works is by
no means trivial and I have no explanation for it".

On the other hand, in Einstein's original derivation,
\cite{EINSTEIN2}, we see the realization that mathematically the
Ricci tensor should be proportional to the source which should be
the total energy density due to both the energy-stress tensors of
matter as well as the gravitational field itself. However, in
\cite{EINSTEIN2}, Einstein was not able to arrive at a fully
covariant tensor expression for the energy density of the
gravitational field. Instead, using a Hamiltonian or variational
method (therefore a weakness in the argument) he arrived at a
pseudo-tensor defined in terms of the connection coefficients for
certain special coordinate systems and which he argued (on grounds
it could be shown that the pseudo-tensor was coordinate divergence
free) served to give the energy density of the gravitational field
for purposes of deriving the equation.

As the arguments in \cite{EINSTEIN1} leading up to the development
of CGR show, Einstein was clearly thinking of the energy of the
gravitational field in a Newtonian way, since in particular, the
connection coefficients are the generalized gravitational forces
from the Newtonian viewpoint.  In particular, in his elementary
analysis of the conversion of gravitational potential to energy
through absorption of a light pulse, he represented the
gravitational potential as height. Moreover, in \cite{EINSTEIN2},
Einstein was very clear that his equation was using the energy
density of the gravitational field in addition to the
energy-stress tensor as the total source of gravity. Indeed, his
derivation there uses a break-up of the already accepted vacuum
equation as an equation for the gravitational field pseudo-tensor
and then he merely argued that the source should have the matter
tensor added in with the pseudo-tensor so that putting the
pseudo-tensor back to the Ricci side of the equation gave the
final fully covariant field equation.

In fact, subsequent attempts to mathematically characterize the
energy of the gravitational field have all basically clung to the
Newtonian framework which makes the energy of the gravitational
field a function of a non-local arrangement of masses and
energies, or combinations of connection coefficients, the results
all giving pseudo-tensors. So much so, that these views are now
taken for granted to the point that in \cite{MTW} we have the
claim of the impossibility of existence of a local energy density
tensor for the gravitational field (see also
\cite{WALD},\cite{HAWKINGELLIS},\cite{FRANKEL},\cite{CW},\cite{SZ}).
This attitude clearly persists to the present as expressed, for
instance, in chapter 3 of \cite{SZ}, or \cite{NEST}.

The non-localizability of gravitational field energy is often, but
unnecessarily, used as a justification for the development of the
profusion of mathematically inspired notions of quasi-local mass,
which all have their advantages and drawbacks as discussed in
\cite{SZ} and \cite{YAU3}, along with extremely involved analysis
required to arrive at their basic properties. In the case of
pseudo-tensors and quasi-local mass definitions developed using
Hamiltonian methods, the results are sensitive to boundary
conditions, as pointed out in \cite{NEST}. As these attempts at
forming quasi-local mass are potentially very valuable for the
global analysis of general relativistic models, the question of
the actual energy density is often irrelevant.

As energy in physics has historically been defined to be a
conserved quantity, the extension of notions of energy in general
relativity have been heavily influenced by this desirable
property. But as soon as Einstein formulated $E=mc^2,$ energy in
relativity took on a physical reality beyond a mere calculation
tool, as it was in Newtonian physics. This means that any
physically real energy, and in particular, gravitational energy,
must be physical, localizable, and itself a source of gravity. For
instance, to quote H. Bondi \cite{BONDI}, "In relativity a
non-localizable form of energy is inadmissible, because any form
of energy contributes to gravitation and so its location can in
principle be found." Whether or not it is conserved then becomes a
separate question. Unfortunately we will find this must be
generally answered negatively, unless the only matter fields
present have energy momentum  stress tensor having zero
contraction, as is the case for the electromagnetic field energy
momentum stress tensor. Trying to force it to be conserved can
only lead to problems. In fact Dirac, the ultimate mathematical
physicist, puts it best, concluding that for the gravitational
field energy, being localizable and being conserved (meaning
divergence free) are not mutually compatible (page 62 in
\cite{D}). Consequently, we must be content to think of the
various pseudo-tensors which are conserved in certain situations
as useful to the extent they are helpful in calculations, but we
should not think of them as giving actual real inertial energy.
Likewise, we must be content to think of the various definitions
of quasi-local mass or energy as tools for calculation and thus
judge them purely on their utility for helping us understand
global solutions to the Einstein equation.

In summary, we give a physical motivation for the postulate that
each observer should view the sum of principal pressures as being
the energy density of the gravitational field he observes.  We
demonstrate that mathematically, Einstein's equation is equivalent
to the combination of three statements (see Theorems
\ref{enrgmmntmstrsgravtheorem} and \ref{theoremeinstein}).  The
first is that each observer should see Newton's Law for
infinitesimal tidal acceleration at his location, in a manner to
be made mathematically precise. The second is that each observer
must include the gravitational energy density he observes in the
source. The third is that each observer must see the energy
density of the gravitational field as the sum of principal
pressures.  The first two statements are obviously the minimal
required modifications of Newton's Law to give a law which makes
sense in relativity, and the third is the postulate which we
motivate physically with an argument involving lasers. The fact
that these assumptions are mathematically equivalent to Einstein's
equation for gravity would seem to make our postulate for the
energy density of the gravitational field very compelling. We
prove a mathematical theorem we call the observer principle which
is really a special case of the uniqueness of power series for
analytic functions which is at the heart of the principle of
analytic continuation. As a consequence of the observer principle,
our postulate that each observer sees the gravitational field as
the sum of principal pressures means that the covariant energy
momentum stress tensor for the gravitational field (see Theorem
\ref{enrgmmntmstrsgravtheorem}) is $T-c(T)g$ where $c(T)$ denotes
the contraction of $T.$

In order to make the presentation clear to a more general audience
than specialists in the field, we have included possibly more
details than an expert will need. Since the dimension of spacetime
does not really enter into the argument, we will actually derive
the gravitation equation for a spacetime of $n+1$ dimensions, and
arrive at the usual Einstein equation in case $n+1=4.$ It is in
higher dimensions that the weakness of pure mathematical arguments
involving desirable forms of equations or of purely
Lagrangian-variational methods becomes clear. It gives the
Einstein tensor as the geometric side of the equation plus other
terms with free parameters \cite{LOVELOCK}. No clearly unique
equation emerges. On the other hand, our energy momentum stress
tensor for the gravitational field dictates a clear choice for the
gravitation equation in higher dimensions. In particular, the
resulting general gravitation equation shows the assumption of
infinitesimal conservation (zero divergence) of the energy
momentum stress tensor of the matter and fields other than gravity
implies that the spacetime must have constant scalar curvature in
spacetime dimension other than $n+1=4.$ This would seem to be a
strong physical indicator that spacetime should be, or at least
appear to be, 4-dimensional. Thus it is only for spacetime of
dimension $n+1=4$ that our derivation gives $div(T)=0$ as an
automatic consequence, where $T$ denotes the energy momentum
stress tensor of all matter and fields other than gravity.

\med

\section{MATHEMATICS AND THE OBSERVER PRINCIPLE}

For general references on differential geometry, semi-Riemannian
and Lorentz geometry, we refer to \cite{LANG}, \cite{KRIELE},
\cite{MTW}, \cite{WALD}, and \cite{ONEIL}. To begin, we assume
that our spacetime is an $(n+1)-$manifold $M$ equipped with a
Lorentz metric tensor $g,$ with signature $(-,+,+,+,...,+),$ and
we denote by $\nabla$ the resulting Levi-Civita Koszul connection
or covariant differentiation operator on $M.$ We use $TM$ for the
tangent bundle of $M$ and $T_mM$ to denote the tangent space of
$M$ at $m \in M.$ If $f: M \lra N$ is a differentiable map of
manifolds, say of class $C^r,$ then $Tf:TM \lra TN$ is the tangent
map which is of class $C^{r-1}$ and we note here the simple
property $T(hf)=(Th)(Tf)$ as regards composition of differentiable
mappings.  In case $f(m)=n \in N,$ then $T_mf:T_mM \lra T_nN$ is a
linear map.  It is convenient in this setting to refer to $u \in
T_mM$ as a unit vector to mean merely $|g(u,u)|=1.$ Thus $u$ is a
time-like unit vector when $g(u,u)=-1.$ Because $M$ is a Lorentz
manifold, each of its tangent spaces is a Lorentz vector space of
dimension $n+1.$

If $u$ is a time-like unit vector in a Lorentz vector space, $L,$
then $u^{\perp} \subset L$ is a Euclidean space. We define the
projection operator $P_u :u^{\perp} \lra u^{\perp}$ by
$P_u(v)=v+g(v,u)u,$ for any $v \in L.$ If $B$ and $C$ are any
linear transformations of $L,$ we say that $C$ is the adjoint of
$B$ to mean that $g(Bv,w)=g(v,Cw)$ for all pairs of vectors $v,w
\in L.$ In this case, $C$ is uniquely determined by $B$ and we
write $C=B^*.$ It is easy to see that in general for any two
linear operators $B$ and $C$ on $L$ we have $(BC)^*=C^*B^*.$ In
particular, $P_u$ is self-adjoint, $P_u^*=P_u$ and as well
$P_u^2=P_u,$ so $P_u$ is an idempotent in the algebra of linear
maps of $L.$

If $B$ is any self-adjoint linear transformation of $L,$ then
$P_uBP_u$ is also self-adjoint but has $u^{\perp}$ as an invariant
subspace and therefore defines a self-adjoint linear
transformation $B_u :u^{\perp} \lra u^{\perp}.$ But since
$u^{\perp}$ is a Euclidean space, this means that $B_u$ is
diagonlizable. We call the eigenvalues (also called proper values)
of $B_u$ the $u-$spatial eigenvalues of $B,$ we call the principal
axes or lines through eigenvectors of $B_u$ the $u-$spatial
principal directions of $B,$ and we call the average of the
eigenvalues of $B_u$ the $u-$isotropic eigenvalue of $B.$ Thus, if
$\lambda_u$ is the $u-$isotropic eigenvalue of $B,$ then
$trace(B_u)=n \lambda_u.$  In particular, if $u$ is also an
eigenvector of $B$ with eigenvalue $r,$ then $B$ is completely
diagonalizable and $trace(B)=r+n\lambda_u.$ But, more generally,
since $g(u,u)=-1,$ we always have,

\begin{equation}\label{trace0}
trace(B)=-g(Bu,u)+n\lambda_u,
\end{equation}
even if $u$ is not an eigenvector of $B.$ Thus, we emphasize that
even though $B_u$ is always diagonalizable, $B$ itself need not
be.

Using the time-like unit vector $u$ allows us to also define a
Euclidean metric or inner product $g_u$ on $L$ by defining
$g_u(v,w)=g(v,w)+2g(v,u)g(u,w).$ This makes $L$ a topological
vector space and in case $L$ is finite dimensional, this gives $L$
its unique vector topology. Thus even though the Euclidean inner
product on $L$ depends on the choice of $u,$ the resulting
topology does not. It is easy to see that for $B=B^*$ to be also
self-adjoint with respect to the Euclidean inner product $g_u,$ it
is necessary and sufficient that $u$ be an eigenvector of $B$ in
which case $B$ is itself then diagonalizable.

If $T$ is a second rank tensor on $L,$ which is merely to say that
$T$ is a real-valued bilinear map on $L,$ then there is a unique
linear map $B_T: L \lra L$ with $T(v,w)=g(Bv,w),$ for all $v,w \in
L.$ We can now invariantly define the {\it contraction} of $T,$
denoted $c(T),$ by

\begin{equation}\label{contract0}
c(T)=trace(B_T).
\end{equation}
Any question of eigenvalues, eigenvectors, or diagonalizbility for
$T$ is really the same question for $B_T.$ Clearly to say $T$ is
symmetric is the same as saying that $B_T$ is self-adjoint. Thus
for $T$ symmetric, its $u-$spatial principal directions are those
of $B_T,$ its $u-$spatial eigenvalues are those of $B_T$ and its
$u-$isotropic eigenvalue is that of $B_T.$ It is customary to call
the $u-$isotropic eigenvalue of $T$ the {\it isotropic pressure}
for the observer with velocity $u$ in case $T$ is an energy
momentum stress tensor and $L=T_mM,$ and we will denote this by
$p_u,$ in this case. Thus for this situation we have, by
(\ref{trace0}) and (\ref{contract0}),

\begin{equation}\label{contract1}
c(T)=trace(B_T)=-T(u,u)+np_u.
\end{equation}
Also, in this situation, $T(u,u)$ is always designated as the {\it
energy density} observed by the observer with velocity $u.$

For $v \in T_mM,$ we denote by $\nabla_v$ the covariant
differentiation operator along $v$ at $m.$ We have then the
Riemann curvature operator, $\R,$ given by

\begin{equation}\label{curvatureoperator}
\R(u,v)=[\nabla_u,\nabla_v]-\nabla_{[u,v]},
\end{equation}
where $u$ and $v$ are any tangent vector fields on an open subset
$U$ of $M.$ We note that $\R(u,v)$ actually defines a vector
bundle map of the tangent bundle $TM|U$ to itself covering the
identity map of $U,$ and it as well then determines the Riemann
curvature tensor, $Riemann,$ of fourth rank, which means that $\R$
is itself an alternating second rank tensor field on $M$ which at
each point $m \in M$ gives a linear transformation valued tensor
on $T_mM.$ In particular, this means that $\R(u,v)$ is defined,
giving a linear transformation of $T_mM$ for any pair of tangent
vectors $u,v \in T_mM.$ One of our main concerns is the certain
contraction of $Riemann$ known as the Ricci tensor, $Ric.$ In fact
in any frame at $m \in M$ with basis $(e_{\alpha})$ for $T_mM$ and
dual basis $(\omega^{\alpha}),$ we have, using the summation
convention,

\begin{equation}\label{ricci1}
Ric(u,v)=\omega^{\alpha}(\R(e_{\alpha},u)v),~~u,v \in T_mM.
\end{equation}
Our notation is chosen to emphasize we are not restricting
ourselves to coordinate frames nor to orthonormal frames unless
explicitly stated. Thus, we will refrain from using the abstract
index notation, as it is too often restricted to imply coordinate
framing. In particular, for any pair of tangent vectors $v,w \in
T_mM,$ the curvature operator defines another linear
transformation $\K(v,w)$ of $T_mM$ defined by

\begin{equation}\label{curv1}
\K(v,w)z=\R(z,v)w,\,\ z \in T_mM.
\end{equation}
Then, among the many basic symmetries of the
curvature tensor, one is immediately equivalent to

\begin{equation}\label{curv2}
\K(v,w)^*=\K(w,v),\, v,w \in T_mM,
\end{equation}
and as

\begin{equation}\label{curv3}
Ric(v,w)=trace [\,\K(v,w)],
\end{equation}
the symmetry of $Ric$ then follows immediately from (\ref{curv2}).
Moreover, if $u$ is any tangent vector, then $\K(u,u)$ is
self-adjoint or symmetric, clearly vanishes on the line through
$u,$ and therefore has $u^{\perp}$ as an invariant subspace. Thus
$\K(u,u)$ really "lives" on $u^{\perp},$ the orthogonal complement
of $u$ in $T_mM.$ We shall denote by $A_u^{(geo)}$ the restriction
of $-\K(u,u)$ to $u^{\perp},$ so $A_u^{(geo)}:u^{\perp} \lra
u^{\perp}$ is a self-adjoint linear transformation of the
Euclidean space $u^{\perp},$ in the case that $u$ is a time-like
unit vector. Thus for this case that $u$ is a time-like unit
vector, the metric tensor is positive definite on this orthogonal
complement, and it follows that $Ric(u,u)$ is simply the sum of
the eigenvalues of $-A_u$ or of $\K(u,u).$ In general, if
$(u,e_1,e_2,...,e_n)$ is an orthonormal frame with $u$ a time-like
unit vector, then we note that

\begin{equation}\label{sectcurv1}
g(e_k,\R(e_k,u)u)=g(e_k,\K(u,u)e_k)
\end{equation}
is the negative of the Riemann sectional curvature of the span of
$u$ and $e_k,$ in $T_mM,$ because $g(u,u)=-1.$ Thus, the
eigenvalues of $A_u^{(geo)}$ are the principal Riemann sectional
curvatures through $u.$  We can now symmetrize and define

\begin{equation}\label{symcurv}
S(v,w)=Sym(\K(v,w)=\frac{1}{2}[\K(v,w)+\K(w,v)],\, v,w \in T_mM.
\end{equation}
We see immediately from (\ref{curv2}) that $S$ is a symmetric
linear transformation valued tensor whose values are themselves
self-adjoint transformations of $T_mM.$  Moreover, we also have
$S(v,v)=\K(v,v)$ for each $v \in T_mM,$ whereas, $Ric(v,w)=trace
\, S(v,w),$ for any $v,w \in T_mM.$

Now, mathematically, $T_mM$ is a Lorentz vector space of dimension
$n+1,$ so as above, taking any time-like unit vector, say $u,$ and
defining $g_u(v,w)=2g(u,v)g(u,w)+g(v,w)$ gives a Euclidean metric
on $T_mM$ making it in particular into a Banach space of finite
dimension. Thus, $T_mM$ is an example of a Banachable space-a
topological vector space whose topology can be defined by a norm.
This topology is actually independent of the choice of $u$ in case
of finite dimensions. Differential geometry can be easily based on
such spaces, and for some examples in infinite dimension, the
interested reader can see \cite{KRIGL&MICHOR}, \cite{BELTITA},
\cite{DUPREGLAZE1} and \cite{DUPREGLAZE2}. In particular, the
theory of analytic functions and power series all goes through for
general Banachable spaces.  We would like to point out how this
can be applied to the theory of Lorentz vector spaces. Suppose
that $E$ and $F$ are Banachable spaces and $S$ is a continuous
symmetric multilinear map (tensor) on $E$ with values in $F,$ of
rank $r.$ We can define the monomial function $f_S:E \lra F$ by
the rule $f_S(x)=S(x,x,x,...,x)=Sx^{(r)},$ and then $f_S$ is an
analytic function. In fact, if $x_1,x_2,x_3,...,x_r \in E,$ then
differentiating, using proposition 3.3 and repeated application of
propositions 3.5 and 3.8 of \cite{LANG}, page 10, we find

\begin{equation}\label{analyticcontinuation}
D_{x_1}D_{x_2}D_{x_3}...D_{x_r}f_S(a)=(n!)S(x_1,x_2,x_3,...,x_r),~a
\in E.
\end{equation}
From (\ref{analyticcontinuation}), we see very generally that if
$U$ is any open subset of $E$ on which $f_S$ is constant, then in
fact, $S=0,$ since we can choose $a \in U.$ Indeed, if $a \in U,$
since $f_S$ is constant on $U,$ it follows that the derivative on
the left side of the equation (\ref{analyticcontinuation}) is 0,
and hence also the right side, for every possible choice of
vectors $x_1,x_2,x_3,...x_r \in E.$ But notice that $a$ does not
appear on the right hand side of (\ref{analyticcontinuation}),
only $S(x_1,x_2,x_3,...x_r),$ and the vectors $x_1,x_2,x_3,...x_r$
can be chosen arbitrarily. Thus, $S=0$ follows. This is just a
very special case of the principle of analytic continuation. We
have therefore proven the following mathematical theorem.

\begin{theorem}\label{obsv1} {\bf ANALYTIC CONTINUATION.}  If $A$ and
$B$ are both symmetric tensors of the same rank, $r,$ on $E$ with
values in $F$ and if $Av^{(r)}=Bv^{(r)}$ for all $v$ in the
nonempty open subset $U$ of $E,$ then $A=B.$
\end{theorem}

We have labelled this as analytic continuation, as it is a well
known special case of the uniqueness of general power series
(there is only one term here). For a purely algebraic proof in the
case $r=2,$ which is the case of most importance here, we refer
the interested reader to \cite{DUPRE}. See also page 72 of
\cite{SACHSWU} for a proof using differentiation for the case
$r=2$ which is similar in form to that given here next.

\begin{corollary}\label{obsv2} {\bf OBSERVER PRINCIPLE.} If $A$
and $B$ are both symmetric tensors of rank $r$ on $T_mM$ with
values in $F,$ and if $Au^{(r)}=Bu^{(r)}$ for every time-like unit
vector in $T_mM,$ then $A=B.$
\end{corollary}

\begin{proof} Since $f_A$ and $f_B$ are homogeneous functions of
degree $r,$ it follows that the hypothesis guarantees
$Av^{(r)}=Bv^{(r)}$ for all $v$ in the light cone of $T_mM$ which
is an open subset of $T_mM.$
\end{proof}

Of course, if we define $U(T_mM)$ to be the set of time-like unit
vectors in $T_mM,$ then this set has a topology called the
relative topology as a subset of $T_mM$ and we have a retraction
function given by normalization which retracts the light cone onto
$U(T_mM).$ It follows immediately that if $W$ is any (relatively)
open subset of $U(T_mM),$ then the hypothesis of the observer
principle can be weakened to merely require $Au^{(r)}=Bu^{(r)}$
for each $u \in W.$ In particular, if we choose a time orientation
on $T_mM,$ then we can merely require $Au^{(r)}=Bu^{(r)}$ for each
future time-like unit vector in $T_mM.$ This is in a sense, the
essence of the {\bf Principle of Relativity}, for instance, as
applied to second rank symmetric tensors-a law (at $m$), say
$A=B,$ should be true for all observers (at $m$) and conversely,
if true for all observers (at $m$), that is if $A(u,u)=B(u,u)$ for
all (future) time-like unit vectors $u \in T_mM,$ then it should
be a law (at $m$) that $A=B.$

We have stated the {\it observer principle} as corollary to the
special case of the mathematical principle of analytic
continuation to emphasize the fact to the casual reader that it is
really a theorem in pure mathematics, and as such, its proof is
completely rigorous.  We call it the observer principle merely to
emphasize how it will be used in what follows.

Notice that the observer principle can be applied to $S$ of
(\ref{symcurv}) as well as to $Ric,$ as tensors on $T_mM.$ Thus,
the observer principle says in a sense that these symmetric
tensors are {\it observable}, in the sense that they are
completely determined at a given event by knowing how all
observers at the event see their monomial forms.

At this point we want to remark that if $E$ is any vector space
with a positive definite inner product, $g,$ and if  $A: E \lra E$
is any linear transformation of $E,$ then $A$ can be viewed as a
vector field on $E,$ say ${\bf v}_A$ where ${\bf v}_A(x)=A(x)$ for
$x \in E,$ and as well it defines the dual 1-form $\lambda_A$ on
the Riemannian manifold $E$ defined by $\lambda_A
(x)(w)=g(A(x),w).$ We record the following result as a proposition
for future use. Its proof is an easy exercise.

\begin{proposition}\label{flatspacedivcurl}
For the linear transformation $A:E \lra E$ of the Euclidean space
$E,$ we have

\begin{equation}\label{flatspacediv}
div_E {\bf v}_A(0)=(div_E A)(0)= trace(A),
\end{equation}
where $div_E$ denotes the ordinary divergence operator on vector
fields defined on $E.$  Moreover, $\lambda_A$ is a closed 1-form
(meaning $d \lambda_A=0$) if and only if $A$ is self-adjoint as a
linear transformation of $E.$
\end{proposition}

For $M$ a Lorentz manifold, $m \in M,$ and $u \in T_mM$ a
time-like unit vector, we will call $u$ an observer at $m.$ We set
$E(u,m)=u^{\perp} \subset T_mM$ and call $E(u,m)$ the observer's
Euclidean space (at $m$). Choose an open subset $W$ of $T_mM,$
with $0 \in W$ and make the choice small enough that the
exponential map carries $W$ diffeomorphically onto a geodesically
convex (\cite{KRIELE}, page 131) open subset $W_L$ of $M$
containing $m.$ Denote the image of $W_E=W \cap E(u,m)$ under this
exponential diffeomorphism by $W_R.$  We shall call $W_R$ the
observer's Riemannian space (at $m$), whereas we refer to $W_E$ as
the observer's Euclidean neighborhood.  Thus, we should
intuitively think of $W_E$ as the Euclidean space an observer
thinks he is in if he is unaware of curvature, whereas $W_R$ is
the space the sophisticated observer thinks he is in when he is
aware of curvature.  Any linear transformation, $A$ of $E(u,m),$
and in particular, $A_u^{(geo)},$ can be viewed by the observer as
a vector field ${\bf w}$ on his Euclidean space, and by
(\ref{flatspacediv}), the divergence, $div_E {\bf w}(0)$ is simply
the trace of $A,$ where $E=E(u,m).$

\med

\section{THE RICCI TENSOR AND SPATIAL DIVERGENCE}

In order to see how the Ricci tensor enters into the theory of
gravity, we should recall the equation of geodesic deviation. If
$[-a,a]$ and $[-b,b]$ is a pair of intervals in $\bR,$ then a
Jacobi field is a smooth map $J:[-a,a] \times [-b,b] \lra M$ such
that for each fixed $\sigma \in [-b,b]$ the map $J_{\sigma} ;
[-a,a] \lra M,$ given by $J_{\sigma}(\tau)=J(\tau,\sigma),$ is a
unit speed geodesic in $M.$  We can then form local vector fields
$u,s$ on an open neighborhood of the image of $J$ in $M,$ denoted
$Im~J,$ so that

\begin{equation}
u(J(\tau,\sigma))=\del_{\tau}J(\tau,\sigma),~~~s(J(\tau,\sigma))=\del_{\sigma}J(\tau,\sigma).
\end{equation}
Thus we must have $[s,u]=0$ and $\nabla_uu=0,$ on $Im~J,$ so we
find

\begin{equation}\label{Jacobi1}
\R(s,u)u=-\nabla_u\nabla_su.
\end{equation}
We will call $s$ in this situation a tangent Jacobi field along
$J_0,$ and at each point it gives the infinitesimal separation
vector. In fact, given $m$ a point on $J_0$ and any unit vector
$s_m \in T_m$ which is orthogonal to $u(m),$ we can arrange that
$s(m)=s_m.$

Since our connection is assumed to be the unique torsion free
metric connection, we have $$[s,u]=\nabla_su-\nabla_us,$$ so the
condition that $[s,u]=0$ gives $\nabla_su=\nabla_us$ in our
present case. In view of (\ref{Jacobi1}), we then find the
equation of geodesic deviation on $Im~J,$

\begin{equation}\label{Jacobi2}
\K(u,u)s+\nabla_u \nabla_u s=\R(s,u)u+\nabla_u \nabla_u s=0.
\end{equation}
In other words, $\nabla_u^2$ "is" the quadratic form of $-\K$ or
$-S$ applied to $u,$ so through $-\K(u(m),u(m))$ we see
$A_{u(m)}^{(ge0)}$ is the linear transformation of $E(u,m)$ giving
the infinitesimal tidal acceleration field, ${\bf
a}_{u(m)}=A_{u(m)}^{(geo)}$ at $m,$ a vector field on $E(u,m)$
defined by ${\bf a}_{u(m)}(s)=A_{u(m)}^{(geo)}s,$ for any
separation vector $s \in E(u,m).$

Of interest to operator theorists here (see \cite{DUPRE0} for
spectral theory and functional calculus of operator fields) could
be the observation that in some sense we have found a relationship
between $\nabla_{u(m)}$ and $(A_{u(m)}^{(geo)})^{1/2}.$

Now, we simply combine the little proposition
(\ref{flatspacedivcurl}) together with (\ref{curv3}), and find at
$m,$ with $0_m$ denoting the zero vector of $T_mM,$

\begin{equation}\label{ricci2}
Ric(u(m),u(m))=trace(\K(u(m),u(m))=-trace(A_{u(m)}^{(geo)})=-div_{E(u,m)}
A_{u(m)}^{(geo)}(0_m).
\end{equation}

We are interpreting this result as relating to the observer's flat
Euclidean space divergence of his flat Euclidean space
infinitesimal tidal acceleration field.  Moreover, by
(\ref{curv2}) $A_{u(m)}^{(geo)}$ is a self-adjoint linear
transformation of the Euclidean space $E(u,m).$

For comparison, we point out that our discussion above for
(\ref{ricci2}) is also the content of results in \cite{ONEIL},
pages 225-219 and 8.9, page 219, as well as \cite{SACHSWU}, 4.2.2,
page 114. We can notice here that by the observer principle,
knowledge of ${\bf a}_{u(m)}=A_{u(m)}^{(geo)}$ for every possible
(even just future pointing) time-like unit vector $u \in T_mM$
would, by (\ref{Jacobi2}) and (\ref{sectcurv1}), determine
$S=Sym(\K)$ at $m$ and thus all Riemann sectional curvatures
which, as is well known in differential geometry (see for instance
\cite{ONEIL}, page 79), in turn determines the entire Riemann
curvature tensor at $m,$ that is both the Ricci curvature and the
contraction free part known as the Weyl curvature (see \cite{WALD}
or \cite{HAWKINGELLIS} for its definition), as pointed out in
\cite{WALD}, pages 41-53.

Because the Weyl curvature is contraction free (all its
contractions are zero), this means the result (\ref{ricci2}) will
only depend on the Ricci tensor and not the Weyl curvature. Thus,
even though the actual tidal acceleration field as a vector field
on $W_R$ would have derivatives in general depending on the Weyl
curvature, the particular combination of derivatives we form, the
observer's flat Euclidean space divergence of the infinitesimal
tidal acceleration, will necessarily be independent of the Weyl
curvature. Of course, if the divergence of the vector field
$\nabla_u^2s$ defined on $W_R,$ the observer's Riemannian space,
is calculated, then the full Riemann tensor enters in and there
seems to be no simple symmetric second rank tensor whose monomial
form will give the Riemannian divergence of the tidal acceleration
field on $W_R,$ and moreover, the Weyl tensor will enter into the
result. Worse yet, as a function of the separation vector at the
given event $m \in M,$ the tidal acceleration in $W_R$ and its
divergence would be a complicated function which would not have
the tensor property as it depends on how the separation vector is
extended to be a vector field in the neighborhood of the event
$m.$

At this point, one might object that the observer could be
rotating which would introduce fictional acceleration into ${\bf
a}_{u(m)},$ and that is correct. A more sophisticated analysis
here could deal with this purely mathematically (see for instance
\cite{FRANKEL}, \cite{MTW}, or \cite{SACHSWU}), but let us allow
that the observer can feel if he is rotating and just say he
restricts to cases where he is not rotating in order to carry out
his measurements. Continuing then, for a non-rotating observer,
the separation or tidal acceleration field in a geometric theory
of gravity is the essence of the gravitational field. That is, if
an observer at event $m \in M$ has velocity $u,$ with $g(u,u)=-1,$
then according to (\ref{Jacobi2}) and (\ref{ricci2}) we should
interpret $R(u,u)$ as the negative flat Euclidean space divergence
of the infinitesimal gravitational tidal acceleration field as
seen by that observer at $m \in M$ who thinks his space is flat
Euclidean. In any case, we shall henceforth simply refer to these
facts as meaning that $Ric(u,u)$ is the Euclidean (space) negative
divergence of the observer's infinitesimal tidal acceleration
field when his velocity is $u \in T_mM.$  Next, we consider how
this relates to Newton's Law of gravity.

\section{NEWTON'S LAW OF INFINITESIMAL \\ TIDAL ACCELERATION}

Let us briefly review how Newton's Law of gravity is formulated
and how it can be recast in terms of the infinitesimal tidal
acceleration field.  Here we have a Euclidean space, $E,$ of
dimension $n$ and a smooth time dependent gravitational vector
field ${\bf f}$ defined on an open subset $U$ of $E,$ where $U$ is
just $E$ with possibly a finite set of points removed which
represent point masses.  Thus the energy density $\rho$ is a
smooth function on $U$ and, in case $n=3,$ Newton's Law says that
$div_E  {\bf f}=-4\pi G \rho$ and ${\bf f}=\nabla \Phi.$  Now this
last condition is easily equivalent to $curl {\bf f}=0,$ since $U$
is simply connected.  On the other hand, keeping in mind that
${\bf f}$ represents an acceleration field (or force per unit
mass), if $s \in T_mE=E$ is a separation vector, then $D_s{\bf
f}(m)$ is the infinitesimal gravitational tidal acceleration with
separation vector $s$ as in \cite{MTW}, pages 272-273 (see also
\cite{OHANIAN&RUFFINI}, pages 38-42). This means that the
infinitesimal gravitational tidal acceleration field is just
$A_m^{(grv)}=T_m{\bf f}$ viewed as a vector field on $E.$ From
(\ref{flatspacedivcurl}) concerning (\ref{flatspacediv}) and the
$curl,$ we see that Newton's Law for gravity in terms of the
infinitesimal gravitational tidal acceleration field $A_m^{(grv)}$
at $m$ says simply $trace(A_m^{(grv)})=-k_n \rho(m)$ and
$A_m^{(grv)}$ is self-adjoint. Of course, this also makes sense
for any spatial dimension $n,$ as in (\ref{flatspacedivcurl}).
Here, $k_n$ is a constant which only depends on the spatial
dimension $n.$ Since $A_u^{(geo)}$ is self-adjoint (\ref{curv2}),
the obvious way to geometrize gravity is simply to identify
$A^{(geo)}$ with $A^{(grv)}.$

Thus on any Lorentz manifold $M,$ we say that the Newton-Einstein
Law for infinitesimal tidal acceleration holds at $m$ for the
observer $u \in T_mM$ provided that $trace(A_u^{(geo)})=-k_n
\rho_u(m),$ where we now view $A_u^{(geo)}$ as the observer's
infinitesimal gravitational tidal acceleration and where
$\rho_u(m)$ is the energy density of matter and fields other than
gravity $u$ observes at the event $m.$ Now, one relativistic
problem with Newton's Law for gravity is the fact that it amounts
to instantaneous action at a distance which conflicts with
relativity. We will assume that this problem is surmounted by {\it
only} requiring the equation to hold at the observer's event $m.$
He can say nothing about events other than his location event as
far as the law of gravity is concerned. This means henceforth, by
definition, and in view of (\ref{ricci2}) that the Newton-Einstein
Law of infinitesimal tidal acceleration at $m \in M$ for the
time-like unit vector $u \in T_mM$ simply states

\begin{equation}\label{newtonlawtides1}
Ric(u,u)=trace(\K(u,u))=-trace(A_u^{(geo)})=k_n \rho_u(m),
\end{equation}
which for short we refer to simply as NEIL at $m$ for observer $u
\in T_mM.$

We say that NEIL holds at $m \in M$ provided that it holds for
each observer $u \in T_mM.$ In this way, we overcome the
relativists claim that there should be no preferred observer. We
say that NEIL holds on $M$ provided that NEIL holds at each point
of $M.$ In this way we make the NEIL into a relativistic universal
law of gravity, which we think of as the Newton-Einstein Law of
gravity. Notice that each observer only claims
(\ref{newtonlawtides1}) to hold at his own event.

We have in fact almost arrived at the correct law, but relativists
could claim we have failed to include all the source energy on the
right hand side of the equation. That is to say, NEIL should be
{\it corrected by requiring that the gravitational field energy
density observed by each observer is also included in the source
term energy density}. We could call this the corrected NEIL or the
CNEIL, but instead we shall call it the Einstein-Hilbert-Newton
Law of infinitesimal tidal acceleration or EHNIL.  Anticipating
the ability of observer $u$ at $m$ to find the energy density of
the gravitational field, $\rho_u^{(grv)}(m),$ we say that the
EHNIL holds at $m$ for observer $u$ provided that

\begin{equation}\label{crrtdnewtnlawtides1}
Ric(u,u)=k_n[\rho_u(m)+\rho_u^{(grv)}(m)].
\end{equation}
Naturally, we then say the EHNIL holds at $m \in M$ provided it
holds for each observer $u \in T_mM,$ and say that the EHNIL holds
for $M$ provided that it holds at each event $m \in M.$  This then
is our {\it universal law of gravitation},  the
Einstein-Hilbert-Newton Infinitesimal Law of Gravity. Of course,
this naturally leads to the question as to the energy density of
the gravitational field which an observer sees at his location
event, which we turn to next.

\section{THE ENERGY DENSITY OF THE GRAVITATIONAL FIELD}

In order to deal with the energy density of the gravitational
field, we must first think in terms of the basic assumption of the
geometric notion of gravity which is that "free test" particles
must follow geodesics. To partially paraphrase J. A. Wheeler,
spacetime tells matter how to move.  That is its job. But
spacetime is the physical manifestation of the gravitational
field, so it is really the job of the gravitational field to tell
matter how to move. Thinking anthropomorphically, if this is the
case, then from the point of view of the gravitational field
itself, it is happiest when all particles are following geodesics.
In fact, we can imagine that in a limiting sense, if "all
particles" follow geodesics, then the gravitational field is
completely relaxed and contains no  energy. It is only when we try
to push a particle off of its geodesic that we feel the reaction
of the gravitational field, and notice we feel it right at the
location of the event of trying to push the particle off of its
geodesic, thinking in the case where $n=3.$

Thus, relativistically, we should think of the manifestation of
tension in the gravitational field is particles not following
geodesics. Now, if a particle is not following a geodesic, then it
is because it is being acted on by a force which is not part of
the gravitational field itself. Because by definition, gravity
acts only through causing particles to follow geodesics, in the
absence of "outside" forces. When a force acts to move a particle
a certain amount off of its geodesic path, the force required to
do so is proportional to the particles inertial mass, by
definition, but in essence, this says the gripping energy of the
gravitational field at the point where the particle is located is
somehow related to the inertial mass of the particle. Accepting
this, the density of this tension energy in the gravitational
field should be related to the force density as manifested in
pressures in various directions.

That is, the energy momentum stress tensor tells us the pressures
as seen by any observer in various directions, so from the energy
momentum stress tensor itself, we should be able to find the
energy density of the gravitational field. For instance, the
pressure you feel on your bottom when sitting in a chair is a
manifestation of the energy density of the gravitational field at
those points on your chair. In a sense then, we could say that if
the surface of your chair were replaced by an infinitesimally thin
slab sitting on top of an infinitesimally lower chair, then the
mass energy of the slab required to hold you in place divided by
the volume of the slab is a reflection of the energy density of
the gravitational field there. What is the minimum mass which can
take care of this job?

In fact, the material the chair is made of in some sense is a
reflection of the energy density of the gravitational field right
where your chair is located. Even primitive people have an
intuitive idea of the strength of material needed to make a chair,
and thus have a working idea of the energy density of the
gravitational field. We should therefore think of the least mass
energy of material required to make a chair as a rough measure of
the energy density of the gravitational field where the chair is
to be used.

More generally, imagine an observer located at $m \in M,$
ghost-like inside a medium with energy momentum stress tensor $T$
and suppose that his velocity at $m$ is $u.$ Then exponentiating
$u^{\perp} \subset T_mM,$ the orthogonal complement of $u$ in
$T_mM,$  the pressures are given by the restriction of $T$ to
$u^{\perp} \times u^{\perp}.$ This is a symmetric tensor on a
Euclidian space so can be diagonalized, as pointed out in the
mathematical preliminaries. Notice that this does not mean $T$
itself is diagonalizable. Thus, there is an orthonormal frame
$(e_1,e_2,...,e_n)$ for $u^{\perp}$ with the property that
$T(e_a,e_b)=p_a \delta^a_b,$ for $a,b \in \{1,2,...,n\}.$ It is
customary to refer to these observed spatial eigenvalues of $T$ as
the principal pressures observed, and their average is referred to
as the observed {\it isotropic pressure}, $p_u.$ Thus, $np_u$ is
the sum of the principal pressures as seen by the observer with
velocity $u.$ Imagine scooping out a tiny infinitesimal box in $M$
at $m$ whose edges are parallel to these $u-$spatial principal
axes of this spatial part of the energy momentum stress tensor. We
can imagine putting infinitesimally thin $(n-1)-$dimensional
reflecting mirrors for walls of the box and filling the box with
laser beams reflecting back and forth in directions parallel to
the edges of the box with enough light pressure in each direction
to balance the force from outside on these reflecting walls.

In a sense, we have standardized a system to balance the pressures
acting to disturb the gravitational field, so we {\it define} the
energy density of the gravitational field as seen by our observer
to be the energy density of the light in this little box. The fact
that a photon has zero rest mass should mean that the light energy
constitutes a minimum amount of energy to accomplish this task of
balancing the gravitational energy. However, it is an elementary
problem in physics to see that the energy density of the light
along a given axis is exactly the pressure in that principal
direction.

Let us review this simple argument, in case $n=3.$ Assume the
coordinates are $(t,x,y,z)$ for simplicity and the box edges are
parallel to these axes with lengths $\delta x, \delta y, \delta
z,$ respectively. Assume that the laser beams parallel to say the
$x-$axis contain $N_x$ photons, each having spatial momentum
$P_x.$ In time $\delta t,$ the photons travel a distance of
$c\delta t$ and hence each such photon makes $(c \delta t)/(\delta
x)$ reflections for a change in momentum of $2P_x$ for each
reflection.

Thus the total momentum transfer to the two end walls
perpendicular to the $x-$axis for the laser beams paralleling the
$x-$axis is

\begin{equation}\label{laser1}
\frac{2P_xN_xc \delta t}{\delta x}.
\end{equation}
This means that the force exerted on the two end walls is
$(2P_xN_xc)/(\delta x).$ But the total area of the two end walls
is $2\delta y \delta z,$ so the pressure on the end walls is

\begin{equation}\label{pressure x}
p_x=\frac{N_xP_xc}{V},
\end{equation}
where $V=\delta x \delta y \delta z$ is the volume of the box. But
the relativistic energy of a photon with momentum $P_x$ is $P_xc.$
Therefore, the total energy density due to the $x-$axis beams is
exactly the pressure in the $x-$direction on the walls
perpendicular to the $x-$axis.

If $p_x$ is negative, a similar argument using opposite charge
distributions on the opposite walls of the box along the
$x-$direction would have the opposite walls behaving like a
capacitor and again, elementary calculations (the freshman physics
"pillbox" argument using Gauss' Law for electric flux) easily lead
to the conclusion that the energy density of the electric field of
the capacitor has the same absolute value as the negative
stretching pressure of the medium, and here it seems that the
energy due to this stretching pressure (like the pressure in a
stretched rubber band) should count as negative energy. Thus, when
pressure is negative, the pressure is serving to reduce the energy
of the gravitational field as it is "working with" the
gravitational field. That is, as we scoop out the matter to create
the little box, if the pressure is negative in the $x-$direction,
we scoop so as to leave opposite charge distributions on the
opposite faces in such a way that the attraction of the opposite
faces balances the negative pressure of the medium. We could
imagine for instance in case the capacitor is overcharged, that
allowing this scooped out capacitor to "snap shut" then supplies
the capacitor energy to the gravitational field and also lowers
the energy of the gravitational field. So in this case of negative
pressure, it must be that the energy density should be negative
just as is the pressure.

We can therefore take it to be the case that the principal
pressure in the $x-$direction gives the energy density
contribution for that direction in any case. Likewise for the
other two axes, consequently we see that the total energy density
in the box is the sum of the pressures that the beams and fields
are balancing, that is the trace of the observer's spatial part of
the energy-stress tensor, $p_x+p_y+p_z.$

More generally, in light of the preceding heuristic arguments, for
any $n,$ we {\it define} the gravitational energy density seen by
the observer $u \in T_mM$ to be the sum of the principal
pressures, $\rho_u^{(grv)}(m)=np_u.$ We now state this as a formal
postulate.

\begin{postulate}\label{postltgrvenrgdnst} {\bf GRAVITATIONAL
ENERGY DENSITY POSTULATE.} At each event $m \in M,$ each observer
$u \in T_mM$ observes the energy density of the gravitational
field as being $\rho_u^{(grv)}(m)=np_u,$ the sum of the principal
pressures of the source matter and fields other than gravity at
event $m.$
\end{postulate}

At this point we can notice that we are already dealing with a
physically intuitive description of the gravitational field which
implies the Cooperstock hypothesis which says the gravitational
field has no energy in the vacuum, because indeed, in the vacuum
there is certainly no pressure. That is obviously postulate
\ref{postltgrvenrgdnst} implies the Cooperstock hypothesis.

Finally here, we should mention that our heuristic argument
involving the laser light box could be replaced by a similar
argument where photons are replaced by any particle which travels
at the speed of light as it then has zero rest mass and therefore
obeys the same energy momentum relation $E=Pc$ as photons do.  For
instance, if we think of the energy of the gravitational field as
residing in particles called gravitons which travel at the speed
of light, and if we assume that gravitons are trying to maintain
geodesic motion of all other matter particles via pressure, then
the same result seems to hold. Thus, maybe $np_u$ is the energy
density of gravitons as seen by the observer $u,$ and thus
gravitons would have no energy in the vacuum, or more precisely,
the vacuum contains no gravitons. Thus, maybe gravitons are the
ultimate constituent particles of matter and fields.

\section{THE ENERGY MOMENTUM STRESS TENSOR\\ OF THE GRAVITATIONAL FIELD}

In view of the results of the preceding section we can now prove
our theorem on the energy density of the gravitational field.

\begin{theorem}\label{enrgmmntmstrsgravtheorem}
If $M$ is a Lorentz manifold and the covariant symmetric tensor
$T$ on $M$ models the energy momentum stress tensor on $M$ due to
all matter and fields other than gravity, then assuming the
Gravitational Energy Density Postulate \ref{postltgrvenrgdnst} is
equivalent to assuming the covariant symmetric tensor

\begin{equation}\label{gravenergydensity}
T_g=T-c(T)g
\end{equation}
is the unique symmetric tensor giving the energy momentum stress
tensor of the gravitational field.
\end{theorem}

For the proof, suppose that $m \in M$ is any event and $u \in
T_mM$ is an observer at $m \in M.$ Suppose that $T$ is the second
rank covariant energy momentum stress tensor for the matter and
fields other than gravity. Our task is to find the covariant
second rank symmetric tensor $T_g,$ which gives the energy
momentum stress of the gravitational field from our previous
physical argument that every observer should see it as the sum of
the principal pressures.  Thus, by the observer principle and the
gravitational energy density postulate \ref{postltgrvenrgdnst},
$T_g$ is uniquely determined by the requirement that
$T_g(u,u)=np_u$ for each time-like unit vector $u \in T_mM$ no
matter which $m \in M.$

Now, applying (\ref{contract1}) to compute $c(T)$  we find that
$$c(T)=-T(u,u)+np_u,$$ and therefore,
$$np_u=T(u,u)+c(T)=T(u,u)-c(T)g(u,u),$$ which is to say finally that
the second rank symmetric covariant tensor $$T_g=T-c(T)g$$ does
indeed do the job.

We point out here, that in general, such uniqueness does not imply
existence, but here we have existence of the required tensor we
seek from equation (\ref{gravenergydensity}) itself. That is
really the assumption that there is a covariant symmetric tensor
$T$ giving the energy momentum stress tensor of all matter and
fields other than gravity is also giving the existence of the
tensor $T_g$ through equation (\ref{gravenergydensity}).

In the reverse direction, by (\ref{contract1}), if we assume that
(\ref{gravenergydensity}) is the energy momentum stress tensor for
the gravitational field, then the gravitational energy density
postulate \ref{postltgrvenrgdnst} is an immediate consequence.
This completes the proof of the theorem
\ref{enrgmmntmstrsgravtheorem}.

In view of (\ref{gravenergydensity}) we define the {\it total
energy momentum stress tensor} of all matter and fields including
gravity to be

\begin{equation}\label{totalenrgmmntmstrss}
H=T+T_g=2T-c(T)g=2[T-(1/2)c(T)g].
\end{equation}
Thus, by Theorem 6.1 and the observer principle, we know $H$ must
be the symmetric tensor which should serve as the source term for
the gravitation equation, since for every $m \in M$ and observer
$u \in T_mM$ we have

\begin{equation}\label{totalenrgdnsty}
H(u,u)=\rho_u(m)+\rho_u^{(grv)}(m)
\end{equation}

\section{THE DERIVATION AND PROOF OF THE EINSTEIN EQUATION}

We are now in a position to state and prove our theorem as regards
the Einstein equation.

\begin{theorem}\label{theoremeinstein}
If $M$ is a Lorentz manifold and $T$ is any covariant symmetric
tensor field on $M$ which models the energy momentum stress tensor
of all matter and fields other than gravity, then the
EHNIL(\ref{crrtdnewtnlawtides1}) together with the Gravitational
Energy Density Postulate \ref{postltgrvenrgdnst} is equivalent to
the assumption that the equations

\begin{equation}\label{einsteinequation2}
 Ric=k_nH=k_n[2T-c(T)g]=2k_n[T-(1/2)c(T)g]
 \end{equation}
 hold on $M$ with $H=(1/k_n)Ric$ being the total energy momentum stress
 tensor of gravity and all matter and fields. In particular, if
 $n=3$ so spacetime is four dimensional, then automatically
 $div(T)=0$ as a consequence of these assumptions.  If $n$ is not
 3, then these assumptions and the assumption that $div(T)=0$
 imply that $dR=0,$ where $R$ is the scalar curvature of $M.$
 \end{theorem}

 To prove the theorem \ref{theoremeinstein} use theorem  \ref{enrgmmntmstrsgravtheorem}. Assuming
 the EHNIL(\ref{crrtdnewtnlawtides1}) holds for all observers
 everywhere,  we now have by (\ref{totalenrgdnsty}),

 \begin{equation}\label{einsteinequation1}
 Ric(u,u)=k_nH(u,u),
 \end{equation}
 for every
 observer $u \in T_mM$ at $m \in M,$
 where $k_n$ is a universal constant depending only on $n.$ As an aside,
 beyond \ref{enrgmmntmstrsgravtheorem} and the EHNIL,
 the real reason behind everything here
 is the fact that $Ric(u,u)$ is the
 negative Euclidean divergence of the tidal acceleration
 (\ref{ricci2}), so in physical terms we are
 using the tracial identification of the gravitational tidal acceleration with the geometric tidal
 acceleration. But let us return to the proof.
 Thus, (\ref{einsteinequation1}) merely
 says that any observer $u$ at any $m \in M$ sees the EHNIL to hold.
 Since (\ref{einsteinequation1}) is
 true for $u$ being any time-like unit
 vector, by the observer principle (corollary \ref{obsv2}), we
 must have (\ref{einsteinequation2}) as an immediate consequence.

 For the reverse direction, if we assume that the equation (\ref{einsteinequation2}) holds
 with $H$ giving the total energy momentum stress tensor of all
 gravity all matter and all fields, then as $T$ is the energy
 momentum stress tensor of all matter and fields other than
 gravity, and as $H=2T-c(T)g,$ we must have $$T_g=H-T=T-c(T)g$$
 which by Theorem \ref{enrgmmntmstrsgravtheorem}
 then implies the gravitational energy density postulate \ref{postltgrvenrgdnst} and then
 (\ref{einsteinequation1}) holds for any observer $u$ which now by (\ref{totalenrgdnsty})
 says the EHNIL(\ref{crrtdnewtnlawtides1}) holds for all observers.

In case $n+1=4,$ it is customary to write $k_3=4\pi G,$ so then

 \begin{equation}\label{einsteinequation3}
 Ric=4\pi G H=8\pi G[T - (1/2)c(T)g],\,\ n=3,
 \end{equation}
 which is a well-known form of Einstein's equation. As $c(g)=n+1$
 and $c(Ric)=R,$ where as usual, $R$ is the scalar curvature, we
 find that $R=(1-n)k_n[c(T)],$ so the equation can be also written
 as $Ric=k_n[2T]+(1/(n-1))Rg,$ and this results in

\begin{equation}\label{einsteineqforgendim=n}
Ric-\frac{1}{n-1}Rg=2k_nT.
\end{equation}

The energy density tensor $T_g=T-c(T)g$ can be expressed in terms
of the Ricci tensor and scalar curvature using
(\ref{einsteineqforgendim=n}) and the result is

\begin{equation}\label{gravengdensgendim=n}
T_g=(\frac{1}{2k_n})[Ric+(\frac{1}{n-1})Rg].
\end{equation}

As usual, we define the {\it Einstein tensor} by
$$Einstein=Ric-(1/2)Rg,$$ which has the property that
$$div(Einstein)=0,$$ no matter the value of $n.$ But in case $n=3,$
we find that the left hand side of (\ref{einsteineqforgendim=n})
is the Einstein tensor. In this case,
 with $k_3=4\pi G,$ we find the most
 familiar form of the Einstein equation

 \begin{equation}\label{einsteinequation4}
 Einstein=Ric-(1/2)Rg=8\pi G~T,\,\ n=3.
 \end{equation}

 Notice that we have not used local conservation of energy,
 $div(T)=0.$  Since the left side of (\ref{einsteinequation4}), the Einstein
 tensor, $Einstein,$ is divergence free, we find $div(T)=0$ as a consequence of our
 derivation, in the case where $n=3.$   On the other hand, it appears that
 for $n$ not equal to 3 we would have that $div(T)$ is in general not zero.  That is, it is only in
 spacetime dimension 4 that the energy momentum stress tensor of matter and
 fields other than gravity can be infinitesimally conserved
 without automatically putting severe restrictions on spacetime. Specifically,
 in case $n+1$ is not equal to 4, we find immediately that $div (T)=0$ implies $dR=0,$ and
 hence the scalar curvature of spacetime must be constant if the energy stress tensor
 of matter and fields other than gravity has zero divergence.

 To see this, in more
 detail, a simple calculation shows that for any smooth scalar
 function $f$ we have $div(fg)=df,$ since $\nabla g=0.$
 The Einstein tensor has vanishing divergence in any dimension
 (due to the second Bianchi identity), and this is clearly
 equivalent to $div(Ric)=(1/2)dR.$  Therefore, taking the
 divergence of both sides of (\ref{einsteineqforgendim=n}) gives
 $$(\frac{1}{2}-\frac{1}{n-1})dR=2k_n div(T).$$  Thus, if we
 assume that $div(T)=0,$ then either $dR=0$ or $n=3,$ and on the
 other hand, if we assume $n=3,$ then as the Einstein tensor has
 vanishing divergence, then so must $T.$ This completes the proof
 of our theorem \ref{theoremeinstein}.

 Before proceeding further, we should remark that it is often
thought that as the Weyl curvature need not vanish in the vacuum,
that it should enter into the expression for $T_g.$ However, we
have a physical expression for the energy density of the field as
seen by any observer, so that determines what the expression for
$T_g$ will be. If the Weyl curvature is not part of the result, we
must accept the fact that Weyl curvature cannot generate
gravitational energy, on this view.

Here, with (\ref{gravengdensgendim=n}), we see explicitly that the
gravitational field energy momentum stress tensor does not depend
on the Weyl curvature. Rather, it only depends on the Ricci
tensor. Thus, our Theorems \ref{enrgmmntmstrsgravtheorem} and
\ref{theoremeinstein}, together with  our gravitational energy
density postulate \ref{postltgrvenrgdnst} in particular guarantees
that the Weyl curvature does not generate gravity. The Einstein
equation determines the Ricci tensor directly from the matter
energy momentum stress tensor.

However, we should keep in mind that including appropriate
boundary conditions, when $n=3,$ the Einstein equation determines
the full Riemann curvature tensor, therefore including the Weyl
curvature tensor. In particular, the reader should note equations
(4.28) and (4.29) on page 85 of \cite{HAWKINGELLIS} for the Weyl
curvature tensor, which follow from the Bianchi identities and are
similar in form to the Maxwell equations for the electromagnetic
field tensor. Thus, the Weyl curvature which gives the curvature
in the vacuum, as the Ricci curvature vanishes in the vacuum, is
contained in the boundary conditions under the Einstein equation.

 We will until further notice now
 restrict to the case $n+1=4,$ so we have ordinary spacetime, and
 therefore $div(T)=0.$

 In the case of $n+1=4,$ if the condition that
 $div(T)=0$ is dropped in the usual derivation where one equates $T$
 to a linear combination of the metric tensor, the Ricci tensor
 and the product of the scalar curvature with the metric tensor,
 and only requires the time components give Newtonian gravity in
 the Newtonian limit, the result is a generalization of the
 Einstein equation with a new free parameter which when equal to 1
 gives the usual Einstein equation.  This has been investigated
 as to its ramifications for cosmology \cite{ALRAWAF}. But, this
 does mean that the assumption $div(T)=0$ is necessary for this type
 of derivation of the Einstein equation, since without it the free
 parameter may be other than unity.

From (\ref{gravengdensgendim=n}) we now have

 \begin{equation}\label{gravenrgydensitygeoform}
 T_g=(1/8\pi G)[Ric+(1/2)Rg]=(1/8\pi G)(Einstein+Rg).
 \end{equation}
 From the last expression on the right, we see, as $T$ and the Einstein
 tensor, $Einstein=Ric-(1/2)Rg$ both have zero divergence, that

 \begin{equation}\label{energydivergence}
 div(H)=div(T_g)=(1/8\pi G)div(Rg)=(1/8\pi G)dR=-d[c(T)].
 \end{equation}
 So even though the total energy and gravitational energy are not
 infinitesimally conserved, the divergence is simply proportional to the exterior
 derivative of  the scalar curvature. Of
 course,  $div(T_g)=-d[c(T)]$ is obvious
 from the definition, (\ref{gravenergydensity}), once we accept $div(T)=0.$
 In particular, as $d^2=0,$ this means that

 \begin{equation}\label{exteriorderivative of grav divergence}
 d[div(H)]=d[div(T_g)]=0,
 \end{equation}
but (\ref{energydivergence}) is even better as it shows $div(T_g)$
is an exact 1-form on $M.$

On the other hand, the equation (\ref{energydivergence}), when
written

\begin{equation}\label{divgravenergy1}
div(T_g)+d[c(T)]=0
\end{equation}
has another interpretation.  In classical continuum mechanics
written in four dimensional form of space plus time, the
divergence of the energy stress tensor equals the density of
external forces.  Of course in relativity, the energy momentum
stress tensor $T$ contains everything and there are no external
forces, as gravity is not a force.  But, we can view
(\ref{divgravenergy1}) as saying that from the point of view of
the gravitational field, the matter and fields represented by $T$
are acting on the gravitational field as an external force density
of $-d[c(T)].$  In classical continuum mechanics, the external
force density has zero time component, but relativistically such
is not the case, the force only has zero time component in the
instantaneous rest frame of the object acted on. We can therefore
view (\ref{divgravenergy1}) as saying that the divergence of the
the gravitational field's energy stress tensor is being balanced
by the rate of increase of $-c(T).$ If $p_x,p_y,p_z$ are the
principal pressures in the frame of an observer with velocity $u,$
where $g(u,u)=-1,$ then $\rho_u=T(u,u)$ is the energy density
observed, and $div(T_g)(u)$ is then the power loss density of the
gravitational field.

Now $c(T)=-\rho_u+p_x+p_y+p_z=-\rho_u+3p_u,$ where $p_u$ is the
isotropic pressure, so (\ref{divgravenergy1}) becomes

\begin{equation}\label{divgravenergy2}
div(T_g)(u)=D_u\rho-3D_up_u.
\end{equation}
Thus, the observer sees the divergence of energy of the
gravitational field is exactly the rate of increase of energy
density of the matter and fields less the rate of increase of
principal pressures.  In particular, in any dust model of the
universe (pressure zero), the gravitational energy dissipation is
exactly balanced by the rate of increase of energy density of the
matter and fields. If $T$ is purely the electromagnetic stress
tensor in a region where there are only electromagnetic fields,
then $c(T)=0,$ and the gravitational energy-stress tensor has zero
divergence, so is then infinitesimally conserved.

To compare our energy momentum stress tensor of the gravitational
field with the various gravitational energy pseudo-tensors, keep
in mind that all examples of such gravitational energy
pseudo-tensors can be made to vanish by appropriate choice of
coordinates and therefore cannot represent real energy of any kind
in relativity.    Such pseudo-tensors generally obey coordinate
conservation laws in appropriately chosen coordinates making them
useful in certain calculations, but they cannot represent real
energy as in relativity, real energy cannot just be transformed
away by some choice of coordinates.  By contrast, our $T_g$ is
fully covariant and represents localizable gravitating energy
density as seen by each observer, but in general is not conserved,
as $div (T_g)$ may not vanish in general.  The exact calculation
of the difference between the various pseudo-tensors and $T_g$ in
various examples should be an interesting problem for future
research.

Looking back at the derivation, one can now see that if there is a
distinction between active gravitational mass and inertial
gravitational mass, then in equation (\ref{crrtdnewtnlawtides1}),
the first of the two terms is the energy density due to active
gravitational mass and the second of the two terms being the sum
of principal pressures is therefore an inertial mass, as it is
inertial mass not following geodesic motion which creates
pressure.  This would mean that in the equation $Ric=4 \pi G
[T+T_g]$ the second term on the right is the tensor which has the
inertial mass whereas the first term is the term with the active
gravitational mass.  But, this would seem to lead to a violation
of the principle of relativity, as the pressures would be
different in different reference frames leading to conversion
between active gravitational and inertial masses depending on the
observer.  Thus, this derivation seems to indicate the equality of
active gravitational mass with inertial mass, a point which is not
addressed in the usual derivations of Einstein's equation.
Possibly an improved version of this derivation might derive the
equality of inertial and active gravitational mass.  Of course,
the geodesic hypothesis itself makes the passive gravitational
mass equal to the inertial mass, which seems to be the reason why
the problem of equality of active gravitational and inertial mass
is often overlooked in elementary treatments of general
relativity.

Finally here, we should point out that Einstein's original
Equivalence Principle is often misconstrued to say that
gravitational fields can be transformed away by choice of
coordinates, and this is certainly not the case, as Frank Tipler
has stated on many occasions.  This is well known to experts in
general relativity. Gravity in general relativity is curvature of
spacetime, and curvature cannot be transformed away. If we view
connection coefficients as "gravitational forces", then using
normal coordinates at a point makes them disappear, but this
merely reflects the fact that gravitational forces do not exist in
general relativity, virtually by definition. Einstein used the
example of an accelerating coordinate system to effectively
transform away a uniform gravitational field in which there is no
actual curvature of spacetime and therefore no real gravity.  At
each event, given a specified limit in level of measurement
accuracy, there is a neighborhood in which curvature effects
cannot then be measured, and in such neighborhoods of an event,
the equivalence principle may be used effectively.  One must be
careful of subtle pitfalls. For instance, when Einstein used the
elevator thought experiment to reason that light would bend in a
gravitational field, he was using the fact that in the accelerated
reference frame the null geodesics appear curved and then
generalizing to arbitrary gravitational fields.

In fact, the elevator thought experiment merely gives the result
for the bending of light that Newtonian gravity in flat Euclidean
space would give under the assumption that photons have inertial
mass. It takes the full Schwarzschild solution to arrive at the
correct answer for the bending of light, which Einstein
fortunately realized before the experimental measurements were
made.

Another way to look at this light bending problem would be that in
NEIL we have out in the near vacuum of space that for the light
beam the law of gravity is $Ric=4 \pi G T_{EM}$ where $T_{EM}$
denotes the energy momentum stress tensor of the electromagnetic
field. But, as $c(T_{EM})=0,$ we have $(T_g)_{EM}=T_{EM},$ so
Einstein's equation, EHNIL, becomes $Ric=4 \pi G [T_{EM}+T_{EM}]=4
\pi G [2T_{EM}]$ which means that the photon's electromagnetic
field gives twice the curvature of the spacetime at points along
its track as would be the case in NEIL, which should reasonably
lead to the doubling of the bending angle. Of course this is a
nonsense argument, since the light bending has to do with tracks
of null geodesics in the gravitational field of a large
gravitating object and not the gravitational field of an
electromagnetic wave itself. But, if we think of a photon passing
a planet, theoretically, we are allowed to think of the planet as
following a path in the gravitational field of the photon, and the
preceding analysis says the planet's path should be bent twice as
much in Einstein's theory as in Newton's theory, so reciprocally,
the photon's track should be bent twice as much. Maybe the
argument is not so specious, and should be examined further. On
the other hand, this does tell us that the effective gravitational
mass of pure electromagnetic radiation or laser light is double
its inertial mass, which possibly could be detected using powerful
lasers in an inertial confinement fusion laboratory, thus leading
to another test of Einstein's theory. Tolman (\cite{TOLMAN},
Chapter VIII) has noticed the prevalence of this doubling effect
for electromagnetic radiation in many examples, all calculated
using the weak field approximation. But, now using Theorems
\ref{enrgmmntmstrsgravtheorem} and \ref{theoremeinstein}, we see
that the EHNIL is telling us the effective gravitational
mass-energy of electromagnetic fields is very generally double the
inertial mass-energy.

More generally, our conclusion here is that $\rho_u+3p_u$ is the
effective gravitational mass-energy density observed by an
observer with velocity $u.$ For that is what is dictated by
Einstein's equation, since it is equivalent to the EHNIL and the
gravitational energy density postulate, by Theorem
\ref{theoremeinstein}.

\section{THE GRAVITATION CONSTANT $G$}

So far, we have not said anything about the determination of the
gravitation constant $G.$  To evaluate this, we merely need to
check the results of experiments with attractive "forces" between
masses. But it is much simpler to just use Newtonian gravity in an
easy example where the results should be obviously approximately
the same.  Consider an observer situated at the center of a
spherical dust cloud of uniform density $\rho,$ and calculate the
tidal or separation acceleration field using Newton's law of
gravitation. We can observe here that the energy momentum stress
tensor satisfies $T(v,w)=\rho g(v,u)g(v,w)$ where $u$ is the
velocity field of the dust cloud.  Thus we calculate easily that
$T_g(u,u)=0$ meaning that a co-moving observer sees the
gravitational field as having energy density zero.  In this case,
the NEIL and EHNIL coincide for $u$ and thus as it seems
reasonable that the NEIL should have the Newton gravitation
constant as its constant, then that means $G=G_N.$

It is easy to give a more elementary argument here. At distance
$r$ from the center, but inside the cloud, the mass acting on test
particles at radial distance $r$ is simply the mass inside that
radius, $M(r),$ by spherical symmetry, as is well-known in
Newtonian gravitation. Here, we have $M(r)=(4/3)\pi r^3 \rho.$

But Newton's Law says the acceleration of a test mass near the
center of the dust cloud is radially inward, and if $r$ is the
distance from the center, then the radial component of
acceleration is given by

\begin{equation}\label{Newton1}
a_r(r)=-G_N\frac{M(r)}{r^2}=-G_N \frac{4 \pi \rho r}{3}.
\end{equation}
Here, $G_N$ is the Newtonian gravitation constant.

On the other hand, considering an angular separation of $\theta,$
the spatial separation is $s=r\theta,$ so the relative
acceleration of nearby test particles in the $s-$ direction
perpendicular to the radial direction is therefore

\begin{equation}\label{Newton2}
a_s(r)=\theta a_r(r)=-G_N\frac{4 \pi \rho r \theta}{3}=-G_N\frac{4
\pi \rho s}{3}.
\end{equation}
Thus the rate of change of separation acceleration of nearby
radially separated test particles in the radial direction at given
$r$ is by (\ref{Newton2}),

\begin{equation}\label{Newton3}
\frac{da_r}{dr}=-G_N \frac{ 4 \pi \rho}{3},
\end{equation}
whereas in the $s$ direction we have the rate of change of
separation acceleration is

\begin{equation}\label{Newton4}
\frac{da_s}{ds}=-G_N \frac{ 4 \pi \rho}{3},
\end{equation}
the same result again.  But there are two orthogonal directions
perpendicular to the radial, so now we see that if ${\bf a}_u$
denotes the spatial or tidal acceleration field around our
observer at the center of the dust cloud, then

\begin{equation}\label{Newton5}
div_u({\bf a}_u)=-G_N 4 \pi \rho.
\end{equation}

As we are dealing with dust, the pressures are zero, so there is
no gravitational energy density, and thus $\rho$ is now the total
energy density seen by our observer.  Thus, we have by
(\ref{ricci2}), that $Ric(u,u)=G_N 4 \pi \rho=4 \pi G_N ~H(u,u).$
But now comparing this result with (\ref{einsteinequation1}), with
$k_3=4 \pi G,$ we see that we must have $G=G_N.$

Notice that in our development, we have used the observer
principle as a form of the principle of general relativity to
reduce everything to working with the time component in an
arbitrary frame for the tangent space. The trick is to be able to
work completely generally so that conclusions apply to $T_{00}$
and $Ric_{00}$ no matter the frame, even in a non-coordinate
frame, which seems best expressed by using $T(u,u)$ and
$Ric(u,u),$ to remind us that we are dealing with an arbitrary
time-like unit vector. It is only now at the end once we have
Einstein's equation that we allow a calculation in a special frame
in order to evaluate the gravitation constant.

Consider for a moment the derivation of Einstein's equation given
in \cite{FRANKEL}. In effect, the derivation of the Einstein
equation given in \cite{FRANKEL} uses the analysis (adapted from
arguments of Tolman \cite{TOLMAN}) of the special case of a static
arrangement of mass for a gravitating fluid drop and adds the
Newtonian gravitational energy density of the fluid drop as
expressed in terms of pressure through the requirement that its
surface pressure be zero to get the time component of the Einstein
equation. Since the setup is a special arrangement of mass, one
cannot assert the observer principle, because the only observer
for which the equation works is the special observer moving with
the drop. However, one can appeal to the general covariance desire
of relativity that equations should be tensor equations valid in
all frames, from which one surmises that if you have found an
equation relating the time components in a special frame, then the
other components in that special frame should also be equal. Once
you accept the full tensor equation in any frame, then it is valid
in all frames and you next surmise that if it works for the liquid
drop, then it must work in general. But, in our present situation,
we have the full equation, in complete generality, and can simply
go backwards through the development in \cite{FRANKEL} to see that
the time component of the equation in the liquid drop case is
Newton's law, and therefore again conclude that our $G$ in
(\ref{einsteinequation4}) is identical to the Newtonian
gravitational constant. For a treatment of linearized Einstein
gravity and its Newtonian approximation in general, one can
consult \cite{MTW} or \cite{WALD}.

At this point, we can simply choose units such that $G=1$ and we
henceforth drop this factor from the equation for simplicity.

\section{THE EINSTEIN DERIVATION}

It is interesting that Einstein realized fairly early in his
search for the gravitation equation that the vacuum equation
should be $Ric=0.$  This lead him to try the equation $Ric=4\pi
G\,\ T,$ as the general gravitation equation when matter is
present. In fact, this equation obviously results from the
observer principle if we assume spacetime satisfies NEIL instead
of EHNIL, that is if our observers neglect the energy density of
the gravitational field. He soon rejected this as not being
compatible with reality, partly due to the fact that it would
require that $div\,Ric=0.$

He also knew that the energy density of the gravitational field
should be included in the source, so if he had found the
expression we have for the energy stress tensor of the
gravitational field, he would have surely arrived at the final
equation at this time.

As it was, in summary, he finally \cite{EINSTEIN2} took the
already accepted vacuum equation $Ric=0$ and for special
coordinate frames, he was able to rewrite the vacuum equation in
the form $s=k(t-(1/2)c(t)g)$ where $t$ is his pseudo-tensor whose
coordinate divergence is zero, and where $s$ itself is a
coordinate divergence of a third rank pseudo-tensor. Let us call a
coordinate system {\it isotropic} provided that in these
coordinates we have $det(g_{\alpha \beta})=-1.$  Thus Einstein
found it useful to restrict to isotropic coordinates.

In fact, he found $Ric=s-k(t-(1/2)c(t)g),$ to be true in any
isotropic coordinate system. He therefore interpreted $t$ as the
energy density of the vacuum gravitational field and interpreted
the new form of the vacuum equation as making $t$ the source.  He
then merely guesses that in the presence of matter with energy
momentum stress tensor $T$ the source should be $t+T$ instead of
merely $t.$ Thus, when $t$ is replaced by $t+T$ in the new form of
the vacuum equation we have $s=k[(T+t)-(1/2)c(T+t)g]$ as the
candidate for the general non-vacuum equation.  We then see that
moving the terms involving the pseudo tensor back to the left side
of the equation results in $Ric=k[T-(1/2)c(T)g],$ true in any
isotropic coordinates.  But this last equation is a fully a
covariant equation.

In a sense, his derivation begins with and is based on the pseudo
tensor for the energy density of the gravitational field.
Technically his equation was $-Ric=k[T-(1/2)c(T)g],$ because he
used a metric with signature $(+---).$ Of course, our summary has
left out the Hamiltonian method he used to arrive at his pseudo
tensor and the considerable technical calculations required to
arrive at the vacuum equation in terms of the pseudo tensor. But,
in outline, it is really quite a nice derivation, and in many ways
superior to most of the modern derivations.

The fact that the coordinate divergence of the pseudo tensor
vanishes means that the general Stokes' theorem (sometimes in this
particular setting called the divergence theorem or Gauss'
theorem) can be applied to give macroscopic conservation of
gravitational energy. On the other hand, once the source $t$ is
replaced by $t+T,$ it is no longer the case that this latter
gravitational pseudo tensor has vanishing coordinate divergence,
so the gravitational pseudo tensor loses its conservation law in
the presence of matter.  It is rather Einstein's total energy
stress pseudo tensor, $T+t,$ material and gravitational, whose
coordinate divergence vanishes. It seems this lead Einstein to
question the need and even the validity for general covariance in
his formulation, since the vanishing of the covariant divergence
could not be integrated to give any macroscopic conservation law.

In our opinion, the real major weakness in the argument is the
reliance on a variational argument using a Hamiltonian to obtain
the pseudo tensor, since there is no apparent way to justify this,
other than picking something that seems simple out of thin air.
Specifically, he chose the integrand to be $g^{\mu
\nu}\Gamma^{\alpha}_{\mu \beta} \Gamma^{\beta}_{\nu \alpha}$ for
his variational integral, where here we can take
$\Gamma^{\alpha}_{\beta
\gamma}=\omega^{\alpha}(\nabla_{e_{\gamma}}e_{\beta}),$ with
$(e_{\alpha})$ the coordinate frame basis and with
$(\omega^{\alpha})$ the corresponding dual frame basis.

For instance, in the Hilbert argument using the scalar curvature
as the integrand, it is certainly simple to write down and after
the fact, it does give the correct equation. But, what is the
physical basis for choosing the scalar curvature for the
variational argument? Without any physical justification, we have
to admit it is just a lucky guess based on trying the simplest
thing, which, of course, is always a good idea when you have
nothing else to go on. Just because you try something simple and
it happens to work does not mean you understand why it works.
After nearly a century of general relativity, we are quite
confident of the results of action principles in general
relativity, but for deriving the equation, it is unsatisfactory.
For instance, the Einstein derivation evolved out of Einstein's
consideration of various physical problems and possibilities and
he happened to arrive at the result at almost the same time as
Hilbert.  Now if Hilbert had proposed his action method two years
earlier, would Einstein have believed the equation was the correct
equation?  Maybe and maybe not. It is putting the cart before the
horse. In fact, setting $Einstein=E,$ if we simply want a simple
derivation, as $E(u,u)$ is half the scalar curvature of $W_R,$ the
exponential Riemannian space orthogonal to $u,$ for any time-like
unit vector $u,$ the simplest derivation is just to guess each
observer sees his spatial curvature proportional to his observed
mass density with a universal constant of proportionality.  This
immediately gives $E(u,u)=2k_3T(u,u),$ for each time-like unit
vector $u$ from which we conclude $E=2k_3T,$ for some constant
$k_3,$ by the observer principle. Instant derivation of the
Einstein equation. But why should we have spatial curvature
proportional to energy density? If you are aware of the observer
principle, it is the obvious guess, but you have no way to know
you are correct, since there is no physics in the argument-it is
just mathematics.

We are not claiming that the Einstein Hilbert Lagrangian method
has no value. It surely has value for certain calculations,
especially since the Lagrangian terms for many fields are known
and can be added in to the calculations. We are simply pointing
out, that if you did not know the equation before such a
derivation, you still might not be convinced.  The derivation we
have presented here seems to have the convincing property that it
is the only way to very generally and naturally "push" Newton's
Law of gravity into a general relativistic framework which
includes the energy density of the gravitational field as seen by
all observers.  For instance, any mathematician familiar with
Newton's Law of Gravity and basic differential geometry would be
convinced by it if he accepts that each observer with velocity $u$
should see $3p_u$ as the energy density of the gravitational
field. It would seem to us that the fact that $div(T)=0$ is an
immediate consequence of this derivation makes it all the more
attractive and convincing.

\section{THE COSMOLOGICAL CONSTANT}

If we include the cosmological constant $\Lambda$ in the Einstein
equation, it becomes

\begin{equation}\label{cosmo1}
Ric -(1/2)Rg+\Lambda g=8\pi T,
\end{equation}
which is of course the same as

\begin{equation}\label{cosmo2}
Ric-(1/2)Rg=8\pi[T-(1/8\pi)\Lambda g],
\end{equation}
which means we view the equation here as having a modified energy
momentum stress tensor

\begin{equation}\label{cosmo3}
T_{\Lambda}=T-(1/8\pi)\Lambda g.
\end{equation}

We then have $c(T_{\Lambda})=c(T)-(1/2\pi)\Lambda,$ so the
effective energy momentum stress tensor of the gravitational field
is

\begin{equation}\label{cosmo4}
(T_{\Lambda})_g=T-c(T)g+(3/8\pi)\Lambda g=T_g+(3/8\pi)\Lambda g,
\end{equation}
and the effective total energy momentum stress tensor serving as
source is

\begin{equation}\label{cosmo5}
H_{\Lambda}=2T-c(T)g+(1/4\pi)\Lambda g=H+(1/4\pi)\Lambda g.
\end{equation}
In any case, as $div~g=0,$ it follows that our conclusions about
the energy-momentum flow of the gravitational field from
(\ref{divgravenergy1}) and (\ref{divgravenergy2}) remain valid,
even in the presence of a cosmological constant.  Equations
(\ref{cosmo4}) and (\ref{cosmo5}) are corrections of equations
(8.4) and (8.5) of \cite{DUPRE} where the numerical coefficients
of the $\Lambda g$ terms were incorrectly given as $1/2 \pi,$ in
both cases.

\section{QUASI LOCAL MASS}

The problem of defining the energy contained in a space-like
hyper-surface has led to many different definitions of the mass
enclosed by a closed space-like surface contained in an arbitrary
spacetime manifold, and these go by the general name quasi-local
mass. Typically, they are defined by some kind of surface integral
and give an indication of the mass enclosed by the space-like
surface. One of the oldest is known as the Tolman integral and is
advocated by Fred Cooperstock \cite{COOPERSTOCK}, \cite{TOLMAN}
(see also \cite{L&L}, equation (100.19), as well as \cite{MITRA1},
\cite{MITRA2}). For an extensive survey of these we refer the
interested reader to \cite{SZ}. In particular, the results of
\cite{TIPLER} on the Penrose quasi-local mass show that the
results can be interesting when the space-like surface is not the
boundary of a space-like hyper-surface.

A list of desirable properties of any definition of quasi-local
mass is given in \cite{YAU3}, where in particular it is shown that
for their definition, the quasi-local mass enclosed by a
space-like surface $S$ is non-negative provided that the dominant
energy condition holds and the surface $S$ is the boundary of a
hyper-surface, $\Omega.$  It is further assumed that the boundary
surface $S$ has positive Gauss curvature and space-like mean
curvature vector, and consists of finitely many connected
components. The local energy condition assumed (equivalent to the
dominant energy condition) is framed in terms of the second
fundamental form of the hyper-surface, and in particular, we can
see that for a geodesic hyper-surface it reduces to the condition
that the scalar curvature of the hyper-surface, $\Omega,$ is
non-negative, since in that case the second fundamental form
vanishes (extrinsic curvature zero). But, in this case, the scalar
curvature of the space-like hyper-surface $\Omega$ is
$2Einstein(u,u)=16 \pi T(u,u),$ where $u$ is a time-like future
pointing unit normal field on $\Omega.$  So if the energy momentum
stress tensor satisfies the weak energy condition in this case,
then the energy density as seen by observers riding the
hyper-surface is non-negative, and we would simply integrate $(1/8
\pi)Einstein(u,u)$ over the hyper-surface to find the energy
inside, which is clearly non-negative.

The amazing result in \cite{YAU3} is that the quasi-local mass
defined there, which is defined in terms of integrals over the
boundary $S,$ is non-negative under the dominant energy condition.
For instance, their results show if the energy inside any one
component of $S$ vanishes, then $S$ is connected and $\Omega$ is
flat (\cite{YAU3}, Theorem 1, page 183), and thus the result shows
that the energy in $\Omega$ is in some sense determined by the
geometry of the boundary and its mean curvature vector under the
assumptions stated above.

The small scale and large scale asymptotic properties are analyzed
in \cite{YU}, and in particular, in the vacuum the result is that
to fifth order the quasi-local mass for small spheres is
asymptotic to the Bel-Robinson tensor whereas in general to third
order it is asymptotic to the energy momentum stress tensor of
matter times volume. Unfortunately, there are drawbacks to this
definition of quasi-local mass, as pointed out in \cite{MURCH},
and it seems the situation is improved with the later treatments
of Wang and Yau, in \cite{WY1}, \cite{WY2}, \cite{WY3}.

Let us use our total energy momentum stress tensor to formulate an
invariant approach to quasi-local energy.  If we have an open
subset $U$ of $M$ and a time-like unit vector field $u$ defined on
$U,$ we can think of the integral curves of $u$ as being the
histories of a field of observers.  We can then form $H(u,u)$ as a
function on $U$ and assuming orientability of $U$ we can choose a
normalized volume form $\mu_U$ so that $\mu_U(u,e_1,e_2,e_3)=1,$
for $(u,e_1,e_2,e_3)$ any local positively oriented orthonormal
frame. The natural way to proceed here seems to be to form a type
of action integral which we can call the {\it mass action
integral}:

\begin{equation}\label{action1}
A(u,U)=\int_U H(u,u) \mu_U=\frac{1}{8 \pi G} \int_U Ric(u,u)
\mu_U.
\end{equation}
The strong energy condition says $Ric(v,v) \geq 0$ for any
time-like vector $v,$ and thus if this condition is satisfied,
then clearly the only way that the action integral can vanish is
for $R(u,u)$ to vanish on $U.$ But, this does not seem to
obviously allow us to conclude that $Ric=0$ on $U.$  However, on
physical grounds, it should allow us to conclude $Ric=0$ on $U.$
That is, if we fill spacetime with observers everywhere, then if
nobody observes any gravitating energy, there should be none.  So
this becomes then a natural mathematical conjecture. In any case,
it would seem that this mass action should be the invariant means
for constructing quasi-local mass.

In order to make use of the total energy-stress tensor, $H,$ in a
setting similar to that of Liu and Yau, \cite{YAU3} or Wang and
Yau \cite{WY1},\cite{WY2}, one would assume an appropriate energy
condition, and then for a space-like hyper-surface $K$ with future
time-like unit normal field $u,$ it is natural to consider
$H(u,u)\mu_K$ where $\mu_K$ is the volume form due to the
Riemannian metric induced on $K.$ The integral of $H(u,u)\mu_K$
over all of $K$ should be the total energy inside $K.$

More generally, if we assume that $H$ is dominantly non-negative,
that is, it satisfies the analogue of the dominant energy
condition for $T,$ then given another reference future pointing
time-like vector field $k,$ one might then integrate $H(u,k)
\mu_K$ over $K.$ If a 2-form $\alpha$ can be found on $K$
satisfying $d \alpha = H(u,k)\mu_K,$ and if $K$ is a 3-submanifold
with boundary $B,$ then by Stokes' theorem, the total energy
inside $K$ is related to the integral of $\alpha$ over the
boundary $B$ of $K.$

In particular, we say that $K$ is {\it instantaneously static} if
there is an open set $U \subset M$ containing $K$ and a vector
field $k$ on $U$ which is future pointing and orthogonal to $K$
and which satisfies Killing's equation, at each point of $K.$ If
$\omega=k^*$ is the dual 1-form to $k,$ so $\omega(v)=g(k,v)$ for
all vectors $v,$ then this is equivalent to requiring $Sym(\nabla
\omega)|K=0$ or equivalently that $(d\omega)|K=2\nabla \omega|K,$
which to be perfectly clear means that the difference
$d\omega-2\nabla \omega$ as calculated on $U$ in fact is zero at
each point of $K.$ Then as in the Komar \cite{KOMAR} integral (see
\cite{PINTONETOSOARES}, \cite{WALD}, pages 287-289 or
\cite{POISSON}, pages 149-151) it follows that

\begin{equation}\label{quasilocalmass0}
(-1/8 \pi)d*d\omega=(1/4 \pi)Ric(u,k)\mu_K=H(u,k)\mu_K.
\end{equation}
Here, $*$ denotes the Hodge star operator on $M.$ Thus, $(-1/8
\pi)*d\omega$ is a potential for the total energy on $K.$ For any
closed 2-submanifold $S$ of $K$ we define the quasi-local total
energy $H(S,k)$ by

\begin{equation}\label{quasilocalmass1}
H(S,k)=-\frac{1}{8 \pi}\int_S *d(k^*).
\end{equation}
Thus, if $K_0 \subset K$ is a submanifold with boundary $S=\del
K_0,$ then by Stokes' Theorem, (\ref{quasilocalmass1}) becomes

\begin{equation}\label{quasilocalmass2}
H(S,k)=-\frac{1}{8 \pi}\int_{K_0} d*d(k^*)=\int_{K_0} H(u,k)
\mu_K,
\end{equation}
which is then non-negative if the strong energy condition holds.
Thus, if $H(S,k)=0,$ with $S=\del K_0,$ then by
(\ref{quasilocalmass2}), under the assumption that the strong
energy condition holds, we would conclude that $Ric(u,k)=0$ on
$K_0.$ But, this means that $Ric(u,u)=0$ on $K_0,$ which means
that none of the observers in the field detect any energy.

Notice that if we have an asymptotically flat spacetime with a
global time-like Killing vector field orthogonal to a spacelike
slice, normalized to be a unit vector at spatial infinity, then
our definition of the quasi-local total energy would be exactly
the Komar mass which is well known in the literature \cite{SZ}.
Thus in the expression $H(S,k),$ the normalization for $k$ is
determined by requiring that it be of unit length at the event at
which the observer is located.  If the observer is located so that
$S$ is in the observer's causal past, then it would seem we must
assume that the domain of $k$ contains this past light cone.

In general, if $k$ is a Killing field on all of the open set $U,$
then being orthogonal to $K$ means (\cite{WALD}, page 119,
(6.1.1)) that also $\omega \wedge d\omega =0,$ where $\omega=k^*.$
Then (see \cite{WALD}, page 443, (C.3.12)) we find, using
$f=ln(|g(k,k)|),$

\begin{equation}\label{quasilocalmass4}
d\omega=-\omega \wedge df,
\end{equation}
and using the fact that here $*[\omega \wedge df]=-(e^{f/2}D_nf)
\mu_S,$ where $n$ is the outward unit normal to $S=\del K_0,$ and
$\mu_S=dA$ is the area 2-form on $S,$ we obtain finally,

\begin{equation}\label{quasilocalmass5}
H(S,k)=-\frac{1}{8 \pi} \int_S e^{f/2}D_nf dA.
\end{equation}
In particular, for the vacuum Schwarzschild solution with mass
parameter $\M,$ taking the Killing field $k=\del_t,$ we see easily
that the mass calculated using the integral
(\ref{quasilocalmass5}) gives the value $\M$ for the mass enclosed
by any sphere centered at the "origin" when we normalize the
Killing field to be a unit vector at infinity.  On the other hand,
if we calculate that value of the integral by normalizing to make
the Killing vector a unit at radial coordinate $r_0,$ as $H(S,k)$
is homogeneous in $k,$ the normalizing constant comes out
resulting in

\begin{equation}\label{quasilocalmass3}
\M_{r_0}=\frac{\M}{[1-\frac{2 \M}{r_0}]^{1/2}}.
\end{equation}
Keeping in mind this is now the total energy, gravitational and
massive, this indicates a problem develops as $r_0 \rightarrow 2
\M,$ even though we know it is not a real problem for the
spacetime. The problem is probably due to the normalization
involving the Schwarzschild radial coordinate which obviously
breaks down at $r_0=2 \M.$ After all, what we are integrating is
equivalent by Stokes' Theorem to integrating $H(u,k)\mu_K,$ when
$S=\del K_0,$ and we really want to be integrating $H(u,u)\mu_K.$
We do not have the actual potential. On the other hand, this does
seem to reflect correctly the fact that as one approaches the
horizon of a black hole it takes infinite force to keep from
falling in.

Let us now use these results to compute the mass action integral
(\ref{action1}). To do this, let us assume that $U$ is foliated by
spacelike submanifolds determined by the Killing parameter $t$ on
$U,$ so the leaves are the level manifolds of $t,$ and that $u$ is
orthogonal to each leaf of this foliation. We assume that $k=hu$
is the Killing vector field on all of $U,$ where $h$ is the
redshift factor. Assume now that $K$ is a compact 4-submanifold of
$U$ with boundary $\del K$ and that the intersection of $K$ with
the leaf at time $t$ is $K_t$ with boundary $\del K_t$ which is
the intersection of $\del K$ with the leaf at time $t,$ for $t_1
\leq t \leq t_2.$ Then $k^*=hu^*$ and we see the volume form on
$K$ can be expressed as $\mu_K=\mu_{K_t}hdt.$ This means that the
action integral can be expressed as

\begin{equation}\label{action2}
A(u,K)=\int_{t_1}^{t_2} \int_{K_t}H(u,u)h
\mu_{K_t}dt=\int_{t_1}^{t_2}\int_{K_t}H(u,k)\mu_{K_t}dt=\int_{t_1}^{t_2}H(K_t,k)dt.
\end{equation}
In the particular case of the Schwarzschild solution, this leads
immediately to

\begin{equation}\label{action3}
A(u,K)=\M \Delta t,
\end{equation}
with the Killing vector normalized so the redshift factor is 1 at
infinity.

We can now see that the real problem is the fact that in
integrating over a spatial slice, the proper time is elapsing at
different rates at different parts of space, so that in general,
the quasi-local mass definitions have to contend with this problem
whether they like it or not \cite{MITRA1}, \cite{MITRA2}.  Thus,
in general, if we have no Killing vector field, if $t$ is an
arbitrary "time" function on $U,$ and if $K(t_1,t_2)$ is the
submanifold of $U$ given by $t_1 \leq t \leq t_2,$ then we should
simply think of $A(u,K(t_1,t_2))=\M_{av}\Delta t,$ where now
$\M_{av}$ is the average quasi-local mass over the given time
interval. This naturally leads to taking

\begin{equation}\label{masst}
\M(t)=\frac{d}{dt}A(u,K(t_1,t)),
\end{equation}
as the mass at time $t.$ Thus for the Schwarzschild solution we
now find that the mass is $\M,$ the mass parameter, which
indicates that the mass parameter is the total mass including that
due to gravitational energy.

Another approach to an invariant treatment of mass in general
relativity might be based upon  the negative of $S$ from
(\ref{symcurv}), and the fact that its restriction $A_u^{geo}$ as
a linear transformation of $u^{\perp} \subset T_mM$ has as its
eigenvalues the negatives of the principal sectional curvatures
which are then the principal tidal accelerations. The eigenvector
of the maximum eigenvalue for an observer at $m \in M$ with
velocity $u$ then picks out a spatial tangential direction in
$u^{\perp}$ which should be either towards or away from any larger
than average matter concentration at locations other than $m.$ For
instance in the Schwarzschild solution, it picks out the radial
direction, and the maximum time-sectional curvature is $2\M/r^3.$
Integrated around a central sphere of radial coordinate $r$ gives
therefore $8 \pi \M/r,$ which is obviously related to the
Newtonian potential of the observer located at radial coordinate
$r$ in the Schwarzschild gravitational field.  This seems to
indicate that there might be a way to obtain a generalization of a
Newtonian type of potential from the curvature tensor in the form
of $-S.$

\section{GRAVITATIONAL RADIATION}

Frank Tipler has pointed out that due to the definition of the
energy momentum stress tensor of the gravitational field in terms
of equation (\ref{gravenergydensity}), it follows that the speed
of sound in the gravitational field equals the speed of light, for
a vacuum electromagnetic field. More specifically, he points out
that the gravitational field energy density tensor equals the
electromagnetic energy density tensor exactly, according to
(\ref{gravenergydensity}) as the contraction of the latter is
zero. Thus the speed of sound in the gravitational field due to a
vacuum electromagnetic field is equal to the speed of sound in a
vacuum electromagnetic field, which is of course the speed of
light. This certainly seems reasonable given our way of viewing
the gravitational energy density in terms of electromagnetic
fields. On the other hand, in the vacuum there is no energy of the
gravitational field, and consequently from this point of view, a
gravitational wave carries no gravitational field energy through
the vacuum.

This point of view has been elaborated previously
\cite{COOPERSTOCK}, in what has become known in the literature as
the Cooperstock hypothesis, purely on mathematical and somewhat
philosophical grounds that the equations for the various pseudo
tensors have no content in the vacuum, and as well, on the basis
of his detailed computation \cite{COOPERSTOCK} involving an
example of a capacitor in a gravitational wave. On the other hand,
the energy density tensor of the gravitational field is not
divergence free which means it can dissipate in one place and
appear in another. That is, the time varying matter tensor causes
gravitational energy to disappear into the vacuum and then
reappear elsewhere where there is matter. This of course is a
difficulty for analyzing gravitational radiation, and it means
that using a coordinate conserved pseudo-tensor or any other
device which is conserved in some useful sense is certainly
justified if it aids in calculation.

For instance, Hayward \cite{HAYWARD}, in analyzing gravitational
radiation in a quasi-spherical approximation defines an energy
density tensor for the gravitational radiation which carries
positive energy and in the second approximation reacts on the
solution when included in the source of the truncated Einstein
equation. This means that one can in special circumstances use
special definitions in a way that can be usefully interpreted
physically, even if it is technically a fiction. On the other
hand, to quote \cite{BORISSOVA}, "At the present time there are
many solutions of the gravitational wave problem, but none of them
are satisfactory...another difficulty: there is no general
covariant d'Alembertian, which being in its clear form, could be
included into the Einstein equations."

We must keep in mind here, that our view of the energy density of
the gravitational field is in complete agreement with the Einstein
equation, so it cannot contradict any of its results and likewise,
no result of solving the Einstein equation can possibly contradict
our view of the energy density of the gravitational field. In
particular, both Carl Brans and Frank Tipler (in personal
communication) have expressed concerns about how the view
expressed here on the gravitational energy momentum stress tensor
relates to the analysis of the energy dissipation from binary
pulsars, an issue also addressed in \cite{COOPERSTOCK} in relation
to the Cooperstock hypothesis. Particularly relevant here are the
calculations in \cite{COOPERSTOCK2} and \cite{COOPERSTOCK3} of the
gravitational radiation due to a rotating rod, showing the general
relativistic calculation to be consistent with the Cooperstock
Hypothesis. The idea that gravitational radiation carries energy
away may be a useful idea for keeping track of the various
"energies", or conserved quantities, in the system, but the
calculations always involve a choice of reference background
metric which produces the apparent "energy". Alternately, it seems
that there is no mathematical vacuum in realistic models of the
universe, because of background radiation and possibly dark
energy, so there is background matter to carry the gravitational
energy. Since the gravitational energy is really $3p_u/c^2$ in
ordinary units, it is so small, that it should be easily carried
by the background matter energy in realistic models involving
ordinary pressures.

\section{BLACK HOLES}

Since the vacuum has no energy density, it follows that the
assignment of mass to black holes or to cosmological solutions is
heavily influenced by boundary conditions assumed for the solution
to the Einstein field equations (see e.g. \cite{NEST}). For
instance, in the case of a Schwarzschild black hole, if we try to
integrate over a region enclosed by a sphere, we find that it is
not the boundary of any compact spacelike slice.  The preceding
analysis leading to (\ref{quasilocalmass3}) would have to be
modified to include also an inner boundary as a cutoff so that the
region bounded is compact.  On the other hand, as it stands, for
the Schwarzschild case, the mass is $\M$ no matter what matter
resides in the interior as long as the matter is not all inside
the Schwarzschild radius, which indicates that it is reasonable to
assign the artificial mass $\M$ to the Schwarzschild black hole
with mass parameter $\M,$ as a reflection of a boundary condition,
the boundary being the black hole horizon.  Thus, in the general
black hole case, one of the various definitions of quasi-local
mass must be adopted. As far as we can see, the actual energy
momentum stress tensor of the gravitational field cannot help
here.

\section{COSMOLOGICAL MODELS}

Because the total energy density is simply $(1/4\pi G)Ric,$ in any
cosmological model where we have a universal time function it is
often straight forward to calculate the total energy density which
thus includes that of the gravitational field.  If the model is
specified by a fluid where $\rho$ is the density and $p$ is
isotropic pressure (the average of the principal pressures)
observed by the universal observer, then $\rho +3p$ is then the
total energy density including that of gravity as seen by the
universal observer. Integrating this over the spatial slice at
time $t,$ if it is compact, gives the total mass of the model
including that due to the gravitational field itself.  In fact, it
has long been realized that $\rho + 3p$ is the actual source of
gravity in general relativity in many special cases and in
particular, in \cite{ELLIS}, we find $\rho + 3p$ referred to as
the "active gravitational mass" of any cosmological fluid model.
This seems to have been clear right from nearly the beginning of
general relativity \cite{TOLMAN}, so it is rather strange that the
energy density tensor of the gravitational field was not realized
right from the Einstein equation itself, as soon as the equation
was accepted. That is, if gravitational energy has itself
effective gravitational mass, then it must be what is accounting
for the extra effective gravitational mass over and above the
ordinary mass.

\section{GENERALIZATIONS}

Our treatment of the Einstein equation depends only on assuming
that there is a Lorentz manifold of dimension $n+1$ which
describes the spacetime model of the universe and that there is a
symmetric tensor field on the spacetime which describes all matter
and fields other than gravity together with the EHNIL at each
point of the spacetime manifold.  However, in our calculations, we
assumed in addition that the constant $k_n$ was a universal
constant, not depending on the particular event $m \in M.$  We can
note that this assumption, though natural, is not implied by the
principle of relativity, so our arguments without this assumption
immediately give the more general equation
$Ric-(1/(n-1))Rg=2k_nfT,$ where $f$ is a smooth function on $M.$
This is because we can still appeal to the principle of relativity
to guarantee that all observers at a particular $m \in M$ would
see the same gravitation constant, which could then depend on $m.$
The observer principle still applies here, as it applies at each
point of $M.$ Using a modification of the Einstein-Hilbert
Lagrangian gives the Einstein equation together with an equation
for the scalar function $f.$ In spacetime dimension 4, this is
usually called Jordan-Brans-Dickie scalar tensor theory
\cite{MTW}. We will keep the constant $k_n$ by giving $f$ the
value 1 at our location. We see that this amounts to replacing the
source $T$ by a new source $fT.$

More generally, Moffat \cite{MOFFAT1} has advocated a scalar
vector tensor theory (SCVT) in order to solve the dark matter
problem. The view in \cite{MOFFAT1} is that SCVT is a new theory
of gravity which solves the dark matter problem (he views the dark
energy problem as solvable with inhomogeneity \cite{MOFFAT2}). But
in fact, a Lagrangian method is used to determine field equations
which can be written in the form $Einstein=8 \pi T,$ where $T$ is
the total energy momentum stress tensor of all ordinary matter and
fields as well as that due to the scalar field and that due to the
vector field and an additional symmetric tensor.  All these extra
structures are required to satisfy equations developed by
Lagrangian methods, and the free parameters can be chosen to match
the dark matter galaxy rotation curves.

However, we can also view these results as being a specific model
for the dark matter within Einstein's theory of gravitation. Thus,
the equations in \cite{MOFFAT1} for the extra scalar and vector
fields of SCVT can be viewed as the beginning of the theory of
dark matter instead of a way of doing away with dark matter. On
the other hand, Frank Tipler \cite{TIPELR2} has argued that such
extreme measures are not needed and that standard physics may be
used to account for the dark matter and energy, as well as several
other problems in cosmology. In any case, it is certainly of
interest that SCVT explains the galaxy rotation curves, explains
the galactic lensing data, explains the bullet cluster data as
well as globular clusters within galaxies, explains oscillations
in the matter power spectrum, all without any additional dark
matter, but fails to explain the pioneer deceleration data as
these two spacecraft are reaching the outer parts of our solar
system. In addition, in SCVT, apparently black holes and
singularities do not exist as there are solutions without event
horizons.  Here, the big bang is replaced by the universe
spontaneously arising from Minkowski space.

In higher dimensions, the usual mathematical method of looking for
the form of the equation of gravity fails to give a unique result,
but rather introduces many free parameters giving a family of
gravitation theories known as Lovelock gravity theories
\cite{LOVELOCK}, \cite{JABBARI}.  A reading of \cite{LOVELOCK}
shows that the theories were discovered without the aid of
Lagrangian methods, but subsequently a Lagrangian was found.  This
shows that without some physics, neither general mathematics nor
Lagrangian methods are capable of arriving at a definitive
equation.  In addition to these theories of gravity, we now have
brane-world theories \cite{RANDALL&SUNDRUM}, \cite{HARKO&CHENG},
\cite{HARKO&MAK}, and $f(R)$ gravity theories \cite {HARKO&b&L}.
Again, these fall into the general scheme of $Einstein=k_nS,$ and
can therefore be rewritten in the form $Ric-(1/(n-1))Rg=2k_nT,$
with $T$ a simple linear combination of $S$ and its contraction
(in case of $f(R)$ we can expand $f$ in power series and the
linear terms inside the Lagrangian give Einstein's equation when
the other resulting terms are moved to the source side of the
equation).

Of course, there are additional fields in these theories which are
required to satisfy additional equations derived by Lagrangian
methods, and which can be thought of as generating new forms of
matter.  In the case of brane-world models for our universe, the
universe we live in is modeled as a Lorentz 4-submanifold of a
higher dimensional Lorentz manifold. In the end, usually effective
equations for the 4-submanifold are found which means again we can
view the result as Einstein's equation with new forms of matter,
as well as a higher dimensional Einstein equation for the bulk.

Thus, in view of Theorem \ref{theoremeinstein}, our equation for
gravity can be viewed as being much more general than in the usual
view.  That is, almost all classical gravity theories are really
systems of equations of which Einstein's equation in form is the
the most important part of the system. Thus we are inclined toward
the view that for any Lorentz manifold model of a universe,
(\ref{gravengdensgendim=n}) should be viewed as the energy
momentum stress tensor of the gravitational field, $(1/2k_n)Ric$
the total energy momentum stress tensor, and
$(1/2k_n)[Ric-(1/(n-1))Rg]$ should be viewed as the geometrically
effective total energy momentum stress tensor of matter and fields
other than gravity.

We can also note here that the Einstein-Hilbert Lagrangian method
gives the Einstein tensor as the geometric side of the equation in
any dimension of spacetime, whereas it is only in spacetime
dimension 4 that the Einstein tensor coincides with our
geometrical tensor $(1/2k_n)[Ric-(1/(n-1))Rg].$ Thus, the Einstein
tensor results from the Lagrangian method in all dimensions
because the coefficient $1/2$ in the term $(1/2)Rg$ results from
the derivative of $\sqrt(-det(g))$ with respect to $det(g)$ which
must be carried out in the variation of the Einstein-Hilbert term
of the Lagrangian, no matter the dimension of spacetime. Moreover,
the fact that conservation of energy makes scalar curvature
constant in higher dimensions indicates that we may be seeing
spacetime as 4 dimensional because that is where everything of
interest is happening, even in higher dimensional theories. As
Norbert Reidel has suggested, maybe the only interesting spacetime
dimensions are 4 and infinity.

\section{CONCLUSION}

The idea that the gravitational field energy can be localized is
not in contradiction of Einstein's equation for gravity, but
rather in fact is a consequence of it.  As soon as we observe the
mathematical fact that the Ricci tensor is giving the negative
flat Euclidean spatial divergence of each observer's spatial
infinitesimal tidal acceleration field, it follows that $(1/4 \pi
G)Ric$ is the total energy momentum stress tensor of matter fields
and gravity, and consequently $T-c(T)g$ must be the energy
momentum stress tensor of the gravitational field. That is, the
Einstein equation is really an infinitesimal law of gravity
governing tidal acceleration which is only a slight correction to
the geometric form of Newton's law for tidal acceleration. We say
merely a slight correction since the correction is only three
times the isotropic pressure, $3p_u$ which in terms of mass
density in terrestrial terms is $3p_u/c^2.$ But these corrections
lead directly and purely mathematically to the Einstein equation
for gravity and the vanishing of the covariant divergence of the
energy momentum stress tensor of all fields other than gravity.

What this means is that the energy momentum stress tensor of the
gravitational field cannot have zero divergence unless spacetime
has constant scalar curvature. This also means that the
gravitational field gets its energy from ordinary matter and
fields other than gravity, confirming the Cooperstock hypothesis
\cite{COOPERSTOCK}. In fact, we are going further than the
Cooperstock hypothesis in that we claim the gravitational field
energy density would even vanish in pressureless dust, for any
observer moving with the dust.  This view seems to be dictated by
the Einstein equation itself. Moreover, if this view is used on
already solved problems and elementary examples and problems, it
should lead to new perspectives.

Additionally, viewed in terms of infinitesimal tidal acceleration,
it seems that each observer sees the effective gravitational mass
density as the inertial mass density plus three times the
isotropic pressure, a possibly testable result.  In particular,
for pure electromagnetic radiation, the effective gravitational
mass density is twice the inertial mass density, which could
possibly be tested in an inertial confinement fusion laboratory.

Finally, our developments are so general as to be able to include
many theories of gravity within the Einstein equation, including
higher dimensional theories. This leads naturally to the point of
view that these theories are really theories involving new forms
of matter within Einstein's theory of gravity.  As well, in case
of higher dimensions, it appears the assumption of the vanishing
divergence of the energy momentum stress tensor for all non
gravitational fields implies that the scalar curvature of
spacetime must be constant.

We can view spacetime itself as the gravitational field, so the
matter is a disturbance of spacetime causing the gravitational
field to have energy momentum stress tensor $T_g=T-c(T)g.$ This
equation is equivalent to $T=T_g-(1/3)c(T_g)g.$ Might not there be
a dual concept of matter field such that the failure of spacetime
to curve so as to be Ricci flat gives the gravitational energy
stress tensor $T_g$ which in turn causes the matter field to have
energy momentum stress tensor $T=T_g-(1/3)c(T_g)g?$ Thus, above we
have attempted to give a physically intuitive way to see how the
energy of the gravitational field arises from ordinary energy. Is
there, dually, a physically intuitive way to view ordinary energy
as arising from gravitational energy?

\section{ACKNOWLEDGEMENTS}

I am very deeply indebted to Frank Tipler for many extremely
useful and helpful conversations and in particular, for making me
aware of the historic problems surrounding the energy density of
the gravitational field, as well as the recent developments in
quasi-local mass. I was unaware of these problems when I developed
these arguments for the energy momentum stress tensor of the
gravitational field and the resulting derivation of the Einstein
equation back in the summer of 2003. I am also indebted to Frank
Tipler for encouraging me to write down the details of these
arguments. I thank Toni Eastham and Juliette Dupre for helpful
conversations and questions.  I would also like to thank Norbert
Riedel, James Glazebrook, and Emma Previato whose comments and
encouragement also lead to improvements in the manuscript. I have
benefited from communications with Fred Cooperstock in the final
stages of preparation of this manuscript, and thank him for
playing Devil's advocate as well as making several very useful
comments. Finally, I would also like to sincerely thank an
anonymous referee whose ideas have lead to several improvements in
the exposition. Of course any errors or controversial views in
what remains are entirely due to the author.

\section{APPENDIX: SYMMETRIES OF THE RIEMANN CURVATURE TENSOR}

We have in (\ref{curv1}) made the definition $$\K(v,w)z=\R(z,v)w$$
so as to define the linear transformation $\K(v,w)$ for any
tangent vectors $v,w \in T_mM,$ and which in particular gives us
the geometric infinitesimal tidal acceleration.  We claimed above
in (\ref{curv2}) that $$\K(v,w)^*=\K(w,v).$$ Here $m$ is a fixed
event in $M.$ Now the basic symmetries of the curvature tensor
\cite{ONEIL} give us

\begin{equation}\label{curve4}
\R(v,w)=-\R(w,v),
\end{equation}

\begin{equation}\label{curve5}
g(\R(v,w)x,y)=g(\R(x,y)v,w),
\end{equation}
and
\begin{equation}\label{curve6}
g(\R(v,w)x,y)=-g(\R(v,w)y,x).
\end{equation}

Obviously, we can see that (\ref{curve6}) is not really
fundamental, as it is an immediate consequence of (\ref{curve4})
and (\ref{curve5}).

So now, using (\ref{curve4}), (\ref{curve5}), and (\ref{curve6}),
we have for any $v,w,x,y \in T_mM,$

$$g(\K(v,w)^*x,y)=g(x,\K(v,w)y)=g(x,\R(y,v)w)=$$

$$g(\R(y,v)w,x)=g(\R(v,y)x,w)=g(\R(x,w)v,y)=g(\K(w,v)x,y).$$

Thus we have shown for any given vectors $v,w \in T_mM$ we have
for all vectors $x,y \in T_mM$ that

$$g(\K(v,w)^*x,y)=g(\K(w,v)x,y).$$ Therefore (\ref{curv2}) holds:

$$\K(v,w)^*=\K(w,v)$$ for all vectors $v,w \in T_mM.$

It is because of (\ref{curv2}) that we immediately find $\K(u,u)$
is self-adjoint with respect to $g,$ the metric tensor.  We thus
seem to be able then to easily make the connection between the
geometric infinitesimal tidal acceleration at $m \in M$ given
through the equation of geodesic deviation and the Newtonian tidal
acceleration given by the derivative of the Newtonian
gravitational force per unit mass vector field at the point $0 \in
T_mM.$

\section{APPENDIX:  GENERAL PRINCIPLE OF ANALYTIC CONTINUATION}

Suppose that $E_1,E_2,...E_n,$ and $F$ are all vector spaces
(possibly infinite dimensional). The function or mapping
$$A: E_1 \times E_2 \times...\times E_n \lra F$$ is a mutilinear
map provided that it is linear in each variable when all others
are held fixed. In this case, we say that $A$ is a multilinear map
of rank $n.$ A useful notation here is just to use juxtaposition
for evaluation of multilinear maps, so we write

$$A(v_1,v_2,...,v_n)=Av_1v_2...v_n$$ whenever $v_k \in E_k$ for $1
\leq k \leq n.$  Thus, we simply treat the multilinear map $A$ as
a sort of generalized coefficient which allows us to multiply
vectors, and the multilinear condition simply becomes the
distributive law of multiplication.

In case that $E_k=E$ for all $k,$ there is really a single vector
space providing the input vectors, and $A:E^n \lra F.$ We say that
$A$ is a multilinear map of rank $n$ on $E$ in this case, even
though in reality, the domain of $A$ is the set $E^n.$ Here it is
useful to write $v^{(k)}$ for the $k-$fold juxtaposition of $v$'s.
Thus we have
$$A(v,v,...,v)=Av^{(n)}.$$ More generally, then for any positive
integer $m$ and vectors $v_1,v_2,...,v_m \in E$ and non-negative
integers $k_1,k_2,...,k_m$ satisfying $k_1+k_2+...+k_m=n,$ we have
the equation
$$Av^{(k_1)}v^{(k_2)}...v^{(k_m)}=A(v_1,...v_1,v_2,...,v_2,...,v_m,...v_m)$$ where
each vector is repeated the appropriate number of times, $v_1$
being repeated $k_1$ times, $v_2$ repeated $k_2$ times and so on.
Of course, if $k_i=0$ then that merely means that $v_i$ is
actually left out, so $v^{(0)}=1$ in effect.

We say that $A:E^n \lra F$ is symmetric if $Av_1v_2...v_n$ is
independent of the ordering of the $n$ input vectors.  Thus when
dealing with algebraic expressions involving symmetric multilinear
maps as coefficients, the commutative law is in effect.

Given any multilinear map $A$ of rank $n$ from $E$ to $F$ we can
define a function $f_A:E \lra F$ by the rule
$$f_A(v)=Av^{(n)}.$$ If we also assume that $A$ is symmetric, then have for any $v_0,v_1,...v_m \in E,$

\begin{equation}\label{multnomthm}
f_A(v_0+v_1+...+v_m)=\Sigma_{[k_0+k_1+...k_m=n]}C(n;k_0,k_1,...,k_m)Av_0^{(k_0)}v_1^{(k_1)}...v_m^{(k_m)}.
\end{equation}
Here $C(n;k_0,k_1,...k_m)$ is the multinomial coefficient:

\begin{equation}\label{multnomcoeff}
C(n;k_0,k_1,...,k_m)=\frac{n!}{k_0!k_1!...k_m!}.
\end{equation}

First suppose that the rank $n$ symmetric multilinear map $A$ has
the property that $f_A:E \lra F$ is constant as a function on $E.$
Let $v_0,v_1,...,v_n \in E.$  Notice we are here dealing with the
case that $m=n.$  There are possibly $n+1$ different vectors here.
Notice that the term on the right-hand side of (\ref{multnomthm})
with $k_0=n$ is simply $Av_0^{(n)}=f_A(v_0).$  Also notice that
the term $k_0=0$ must have $k_1=k_2=...=k_n=1,$ which gives
$n!Av_1...v_n.$  Let $w=v_0+v_1+...+v_n.$  Since we assume that
$f_A$ is constant on all of $E,$ it follows that $f_A(w)=f_A(v_0)$
and therefore by (\ref{multnomthm}), for all vectors
$v_0,v_1,...,v_n \in E,$ we have

\begin{equation}\label{multi2}
n!Av_1v_2...v_n+\Sigma_{[k_0+k_1+...k_m=n, 0 < k_0 <
n]}C(n;k_0,k_1,...,k_m)Av_0^{(k_0)}v_1^{(k_1)}...v_m^{(k_m)}=0.
\end{equation}
Notice that $v_0$ can now be taken equal to 0 in equation
(\ref{multi2}) and the result is $n!Av_1v_2...v_n=0,$ for any
vectors $v_1,v_2,...,v_n \in E.$  This means that $A=0.$  We have
therefor proven

\begin{proposition}\label{genanalcont} If $A$ is a symmetric multilinear map on a
vector space $E$ with values in the vector space $F$ and if the
function $f_A:E \lra F$ is constant on $E,$ then $A=0.$
\end{proposition}

Suppose now that $E$ is a topological vector space and that $U$ is
a non-empty open subset of $E$ on which $f_A$ is constant.  Let
$v_0 \in U$ and let $w_1,w_2,...,w_n \in E$ be arbitrary.  Because
$U$ is open in $E$ and the operations of vector addition and
scalar multiplication are continuous mappings, it follows that
there is a positive number $\epsilon$ so that if $t_1,t_2,...,t_n$
are any numbers with $|t_i| \leq \epsilon,\,1 \leq i \leq n,$ then

$$v_0+t_1w_1+t_2w_2+...t_nw_n \in U.$$
If we now take $v_i=t_iw_i$ for $1 \leq i \leq n$ in our previous
calculation (\ref{multi2}), we find that if $|t_i| \leq
\epsilon,\,1 \leq i \leq n,$ then

\begin{equation}\label{multi3}
n!(t_1t_2...t_n)Av_1v_2...v_n+\Sigma_{[k_0+k_1+...k_m=n, 0 < k_0 <
n]}(t_1^{(k_1)}...t_n^{(k_n)})C(n;k_0,k_1,...,k_m)Av_0^{(k_0)}w_1^{(k_1)}...w_m^{(k_m)}=0.
\end{equation}
In particular, this means that

\begin{equation}\label{multi4}
n!\epsilon^nAv_1v_2...v_n+\Sigma_{[k_0+k_1+...k_m=n, 0 < k_0 <
n]}\epsilon^nC(n;k_0,k_1,...,k_m)Av_0^{(k_0)}w_1^{(k_1)}...w_m^{(k_m)}=0.
\end{equation}
and therefore

\begin{equation}\label{multi4}
n!Av_1v_2...v_n+\Sigma_{[k_0+k_1+...k_m=n, 0 < k_0 <
n]}C(n;k_0,k_1,...,k_m)Av_0^{(k_0)}w_1^{(k_1)}...w_m^{(k_m)}=0.
\end{equation}
But then, by (\ref{multnomthm}) and (\ref{multi4}), we have
$f_A(v_0)=f_A(v_0+w_1+w_2+...+w_n)$ no matter the choice of
vectors $w_1,...,w_n \in E.$  This means that $f_A$ is constant on
$E$ and by Proposition \ref{genanalcont} we now conclude that
$A=0.$  We have now proven

\begin{proposition}\label{gentopanalcont}
suppose $E$ is any topological vector space and $F$ is any vector
space. Suppose that $A:E^n \lra F$ is any symmetric multilinear
map of rank $n.$ If there is a non-empty open subset of $E$ on
which $f_A:E \lra F$ is constant, then $A=0.$
\end{proposition}

By convention, a multilinear map from $E$ to $F$ of rank zero is
just a vector in $F.$  If $A_k$ is a symmetric multilinear map of
$E$ to $F$ of rank $k,$ for $0 \leq k \leq n,$ then the function

$$f=\Sigma_{k=0}^n f_{A_k}$$
is a polynomial function of degree $n.$ If $F$ is also a
topological vector space, then we can take limits in the sum and
consider power series. The general principle of analytic
continuation relies on the uniqueness of power series expressions.
In general, for Banach spaces, if two power series agree locally
as functions, then all their coefficients are the same-that is,
they are the same power series. The proof is easy using
differentiation, just use the same method used in freshman
calculus, but for Banach space valued functions. We have basically
proven this fact in case there is only one term in the power
series, but without using differentiation and without even having
topology on the range vector space.

The Observer Principle as we have formulated it here is just this
special case of the principle of analytic continuation given in
Proposition \ref{gentopanalcont}.  In a sense, it is the essence
of the Principle of Relativity.  Because it says that in order for
two rank $r$ symmetric tensors $A$ and $B$ on $T_mM$ to agree,
$A=B,$ it merely suffices that $Au^{(r)}=Bu^{(r)}$ for every
observer $u \in T_mM.$ A law in general relativity at event $m \in
M$ expressed as an equation of symmetric tensors is true if and
only if each observer sees the two tensors as equal.  We can think
of observer $u \in T_mM$ observing the rank $r$ tensor $A$ on
$T_mM$ by actually finding the value of $Au^{(r)}.$  For further
discussion of this topic, see the Appendix of \cite{DUPRE}.

\section{APPENDIX: COMMENTS OF ENERGY DENSITY OF GRAVITY}

We must keep in mind here that our "scooping out" in either the
positive pressure case or the negative pressure case is really
just a thought experiment in the sense of one of Einstein's
"gedanken" experiments.  We are just imagining that for an
infinitesimal amount of time $\delta t$ as seen by observer $u,$
that as if by magic, a small part of physical reality were
replaced by either laser beams or capacitor as the case may be.
Thus, in the negative energy case, we are not asking the observer
to sort the charges along the cut line, we are merely asking the
reader to imagine what the observer would see
"energy-density-wise", if for a certain infinitesimal duration
$\delta t$ his physical reality was magically modified with
electromagnetic fields so as to balance the pressures.  Notice,
that for the laser beam argument, we must assume that the duration
is large enough that statistically many photons strike the
opposite faces for the creation of photon pressure.  This means
that $c \delta t$ should be large in comparison to $\delta x,
\delta y, \delta z,$ but as we are dealing with infinitesimals in
the macroscopic sense of physics, this is not a problem.

For more detail in the "pillbox" argument, we must assume that the
charge surfaces behave like conductors so that there is no field
outside the scooped out region nor parallel to the scooped regions
opposite pair of faces which we are considering. This, of course,
is somewhat of a stretch. We must assume that the capacitor's
electric field makes no change to the region outside. Suppose
${\bf E}$ is the electric field vector inside the scooped spatial
region and that $\sigma$ is the surface charge density. By Gauss'
Law for electric flux, we then have the electric field flux $|{\bf
E}|A=A \sigma,$ where $A$ is the surface area of one of the faces
of the scooped out region across which we have the negative
pressure, say $A=\delta y \delta z,$ and the separation of two
charged surfaces is $\delta x.$ Thus,
$$|{\bf E}|=\sigma.$$ The energy of two such separated charged
surfaces is therefore $|{\bf F}| \delta x$ where ${\bf F}$ is the
force exerted by one face on the other.  But then ${\bf F}=A
\sigma {\bf E}$ so the energy is
$$|{\bf F}| \delta x=\sigma |{\bf
E}|A \delta x= A \sigma^2 \delta x=\sigma^2 V,$$ where $V=A \delta
x$ is the volume.  Thus the energy density equals the square of
the surface charge density. On the other hand, the pressure in the
$x-$direction here in absolute value satisfies $$A|p_x|=|{\bf
F}|=A \sigma |{\bf E}|=A \sigma^2.$$ Therefore, the absolute value
of the pressure also equals the square of the surface charge
density.  Thus, we conclude that the energy density of an electric
field inside the scooped region caused by charges on opposite
faces attracting each other is exactly the pressure in absolute
value.

There are clearly problems with the capacitor argument as far as
actually physically putting it into effect is concerned.  For
instance, the edge effect of a finite parallel plate capacitor
which causes electric field lines to "bulge out" is certainly
undesirable, and the argument would be ruined by the electric
fields created outside the scooped out region.  Somehow, the fact
that the energy density equals the pressure in both the positive
and negative pressure cases here, however, seems to be too much of
a coincidence to ignore.  This seems to indicate that there must
be some very strong connection between gravity and
electromagnetism.  For instance, one might imagine that a whole
space-like slice of spacetime is chopped up into such tiny
infinitesimal bits and everything replaced with such infinitesimal
electromagnetic field systems.  How would the observers know?  All
the pressures they feel would be the same everywhere.  I would
love to hear comments from physicists on this.

\end{document}